\pdfoutput=1 
\documentclass[sigconf,10pt]{acmart}
\usepackage{hyperref}
\usepackage{breqn,xspace}
\usepackage{subfigure, bm}
\usepackage{multirow}
\usepackage{threeparttable}
\usepackage{mathrsfs}
\usepackage{pifont}
\usepackage{caption}
\usepackage{subfigure}
\usepackage[noend]{algorithm2e}
\usepackage{dblfloatfix}
\usepackage{lipsum,multicol}

\AtBeginDocument{%
	\providecommand\BibTeX{{%
			\normalfont B\kern-0.5em{\scshape i\kern-0.25em b}\kern-0.8em\TeX}}}
\acmConference[Preprint copy]{The 30th Annual International Conference on Mobile Computing and Networking}{UW-Madison.}{USA}
\acmBooktitle{The 30th Annual International Conference on Mobile Computing and Networking (ACM MobiCom'24), September 30 -- October 4, 2023, Washington, D.C., USA}
\renewcommand\footnotetextcopyrightpermission[1]{} % removes footnote with conference info
\setcopyright{none}

\newcommand{\mm}{mmWave\xspace}
\newcommand{\name}{Gemini\xspace}

\ifodd 1
\else

\fi

\begin{document}

%\title{\name: Integrating Monostatic Sensing with Communication upon Millimeter Wave Radio}
\title{\name: Integrating Full-fledged Sensing upon Millimeter Wave Communications}
\author{\rm Yilong Li$^{1}$, Zhe Chen$^{2}$, Jun Luo$^{3}$, Suman Banerjee$^{1}$}
\affiliation{University of Wisconsin-Madison$^{1}$ \country{USA}, Fudan University$^{2}$ \country{China}, \\
            Nanyang Technological University$^{3}$ \country{Singapore}}	

\begin{abstract}
Integrating millimeter wave (mmWave) technology in both communication and sensing is promising as it enables the reuse of existing spectrum and infrastructure without draining resources.
Most existing systems piggyback sensing onto conventional communication modes without fully exploiting the potential of integrated sensing and communication (ISAC) in mmWave radios~(not \textit{full-fledged}).  
In this paper, we design and implement a \textit{full-fledged} mmWave ISAC system \name; it delivers raw channel states to serve a broad category of sensing applications.
%to bridge the gap between state-of-the-art communication architecture and the new ISAC one. 
%
We first propose the mmWave self-interference cancellation approach to extract the weak reflected signals for near-field sensing purposes. 
% and then employ an off-the-shelf 60\!~GHz mmWave radio front-ends with 40-element phased arrays to further improve those signals. 
Then, we develop a joint optimization scheduling framework that can be utilized in accurate radar sensing while maximizing the communication throughput. Finally, we design a united fusion sensing algorithm to offer a better sensing performance via combining monostatic and bistatic modes. We evaluate our system in extensive experiments to demonstrate \name's capability of simultaneously operating sensing and communication, enabling mmWave ISAC to perform better than the commercial off-the-shelf mmWave radar for 5G cellular networks.
\end{abstract}
\maketitle

%\keywords{Integrated sensing and communication (ISAC), mmWave communications, monostatic sensing, deep learning.}

\vspace{-1em}
\section{Introduction}
\label{sec:intro}
\vspace{-0.5ex}
% The past decade has witnessed significant advances in using RF signals for Sensing and communication individually. Novel algorithms and applications have been studied and developed in wireless community. Millimeter wave (mmWave) is playing an important role in next generation wireless LANs, and IoT devices for various sensing and communication purposes. It allows us to use narrow beams and shield a mmWave device from interference outside the main direction (main lobe) of its beam. Also, numerous applications have been proposed to utilize mmWave sensing, i.e., vehicular imaging and medical sensing. Other mainstream perception modalities, i.e., camera and Lidar, are vulnerable under challenging lighting and weather conditions.

The past decade has witnessed significant progress in millimeter wave (mmWave) technology since it offers wider bandwidth and more antennas (e.g., 2\!~GHz bandwidths and 16-element antennas~\cite{mmFLEX-MobiSys20}), compared with commodity sub-10\!~GHz technologies (e.g., Wi-Fi 6~\cite{802_11ax,wifi6}). On the one hand, these features enable mmWave communications to support a wide range of high-throughput applications, such as ultra HD video streaming, virtual and augmented reality~\cite{MoVR-NSDI17, TengVR-MobiCom17}. On the other hand, the same features also endow mmWave sensing with high spatial resolution, making mmWave radars necessary components of 
% have been widely deployed in the current 
smart vehicles and robots~\cite{mmV2X-MobiCom20, milliMap-MobiSys20, radatron-ECCV22}. 
%Whereas these two trends used to be developed independently, 
While progress in communication and sensing used to be fairly independent,
both academia and industry have started exploring the promising \textit{integration of sensing and communication}~(ISAC) for the next generation of mmWave systems~\cite{SPARCS-IPSN22, liu2018toward}, enabling the use of existing spectrum and infrastructure, avoiding the cost of new hardware.

% Although 5G millimeter Wave (mmWave) boosted networks globally and delivered multi-gigabit speeds, capacity and exceptionally mobile broadband speeds in suburban and rural communities, it needs dense deployment and increases hardware cost a lot. Nowadays, integrated Sensing and Communication (ISAC) has attracted substantial attraction in recent years for spectral efficiency improvement, enabling hardware and spectrum sharing for simultaneous sensing and signaling operations. It is beneficial to have a joint communication and radar system that allows hardware reuse.

% The Augmented Reality (AR) / Virtual Reality (VR) technique has grown rapidly and Metaverse concept gained a lot of attention, cost-sensitive and low power mobile network connections play a significant role in such products. For example,
A number of existing theoretical proposals on \mm ISAC systems for 6G and beyond have appeared~\cite{fan6GISAC-JSAC22} focus mainly on focus on waveform design, rather than guiding the practical systems to merge one function (e.g., sensing) with devices designed for another purpose (e.g., mmWave Wi-Fi~\cite{mmtrack-INFOCOM20, mmeye-IoTJ}).
%
%Though numerous theoretical proposals on mmWave ISAC systems for 6G and beyond have appeared~\cite{fan6GISAC-JSAC22}, they often focus on waveform design, rather than guiding the practical systems to merge one function (e.g., sensing) with devices designed for another purpose (e.g., mmWave Wi-Fi~\cite{mmtrack-INFOCOM20, mmeye-IoTJ}).
% these studies are limited by the practical implementation of such systems. 
Among the few existing mmWave ISAC systems, most are designed to piggyback sensing onto communication infrastructure, hence confining sensing to only the \textit{multi-static} communication setting with transmitter (Tx) and receiver (Rx) physically separated~\cite{SPARCS-IPSN22}.
% a communication-oriented approach, such as bistatic or even multi-static sensing, but sacrifice communication performance. 
Only one SDR-based system by far has emulated a radar-like \textit{monostatic sensing} capability~\cite{guan-TMTT2021}, where the Tx and Rx are co-located to enable precise range estimation, yet it bears no intention in realizing ISAC: its
% co-locates a pair of Tx-Rx antenna arrays, rendering it a 
waveform is designed only for sensing, rendering it
% prototype 
% not suitable for indoor scenarios and not 
largely incompatible with the existing commodity mmWave devices~\cite{intelwigig, qualcommwigig}. In a nutshell, while theoretical studies on ISAC barely contribute to the development of real systems, existing system design stays at \textit{adapting} phase far from a full \textit{integration}.
% we need to design a new and revolutionary system to realize the ambition of mmWave ISAC. 

% The study presented in~\cite{} claims to enable far-field monostatic sensing with a bandwidth of only 160\!~MHz for RF imaging applications at 28\!~GHz. However, this approach does not provide short-range monostatic sensing capability and  the bandwidth is still too narrow for mmWave. revised[Therefore, there is a need for further research to fully explore the potential of mmWave ISAC systems. ]
% protocol 
\begin{figure}[t]
\vspace{-1ex}
  \setlength\abovecaptionskip{8pt}
  \centering
  \includegraphics[width=.86\columnwidth]{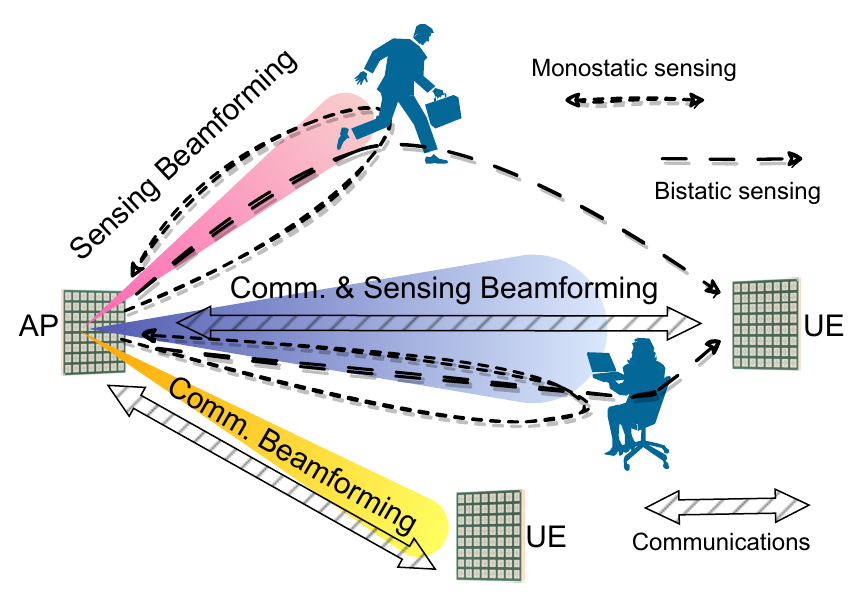}
  % \vspace{-0.3cm}
  \caption{Vision of full-fledged mmWave ISAC system.}
  \label{fig:teaser}
  \vspace{-1em}
\end{figure}

% For a typical mmWave ISAC scenario shown in Figure~\ref{fig:teaser}, there are three types of beamforming including sensing, communication and communication as well as sensing simultaneously. 

Unlike conventional sub-10\!~GHz Wi-Fi often getting only a couple of antennas, mmWave communication systems have their access point (AP) and user equipment (UE) both equipped with multiple phased arrays~(a.k.a. hybrid beamforming for massive MIMO~\cite{RobertsFDmmWave,ISAChybridbeamforming-ICC22}). As a result, typical mmWave ISAC scenarios shown in Figure~\ref{fig:teaser} have their communication links highly directional, 
making them sub-optimal for the multi-static sensing setting where sensing subjects are often off the link.
%making them at odds with the multi-static sensing function whose sensing subjects are often off the link.
%
% mmWave can leverage them to provide three different types of beamforming. Moreover, from the wireless sensing links perspective, the sensing modes are divided into monostatic~\cite{} and bistatic~\cite{}. Apparently, most of current mmWave ISAC studies~\cite{} simplify the complicated ISAC cases to the communication and sensing beamforming simultaneously with bistatic mode case that the sensing target and UE are in the similar direction, but are not joint communication and sensing optimization work for all cases. Consequently, those systems are still far from the full-fledged mmWave ISAC system. 
%
Therefore, a full-fledged mmWave ISAC system would need to better leverage the beamforming capability to handle sensing and communication in a truly integrated manner. In particular, the beam patterns should be designed to support three functions: i) pure communication, ii) pure sensing under monostatic (reflection) sensing mode, and iii) simultaneous communication and sensing under monostatic and multi-static modes, as illustrated in Figure~\ref{fig:teaser}.
%
% The conventional communication-oriented design of mmWave ISAC systems only considers the case of communication and sensing beamforming simultaneously and the bistatic sensing mode that are the subset of full mmWave ISAC. Apparently, the full-fledged mmWave ISAC system is very complicated, and hard to achieve.    

% To design and implement a perfect mmWave integrated sensing and communication system, we need to ask a question: can we achieve a better isolation to separate the communication signal and sensing signal?

Implementing such a truly integrated mmWave ISAC system faces three major challenges. Firstly, unlike long-range sensing where direct Tx interference can be readily removed thanks to the fine-grained temporal resolution resulting from mmWave's wide bandwidth, 
% is too weak to submerge the reflection signals from the target. But for 
short-range sensing commonly adopted for indoor scenarios has to cope with the Tx interference potentially overwhelming the reflected sensing signals~\cite{isacot}. 
Second, though beam scheduling exists for mmWave communications, satisfying the three functions of mmWave ISAC demands a largely enhanced fair scheduling algorithm to weigh among all necessary beam patterns.
Last but not least, both monostatic and multi-static modes may coexist and thus generate complementary sensing information concerning the same subject, incurring the need for a unified framework to fuse such diversified information in a constructive manner.

% Aiming to address the above challenges, 
To this end, we build \name as a system that effectively integrates full-fledged sensing and communication at mmWave band.
%
% We first theoretically and mathematically model the  fully-fledged mmWave ISAC system design challenges as the optimization problem. Then, since that optimization problem is non-convex, We propose three strategies to solve it: i) We desing on-linear interference cased by direct and cross-talk Tx signals.
%
\name employs both a smart beamforming and a deep neural model to cancel the 2\!~GHz wideband Tx interference; this allows the weak reflected signals from sensing subjects to be effectively extracted for short-range sensing. Extending the idea of smart beamforming, \name further innovates in a set cover inspired beam scheduling algorithm, in order to satisfy both sensing accuracy and communication throughput. Finally, \name is equipped with a distributed fusion mechanism for unified estimation, leveraging diversified information gathered from both monostatic and multi-static sensing modes. 
% Finally, all these novel mechanisms 
All these are wrapped into a mmWave ISAC protocol largely compatible with existing 802.11ay~\cite{802_11ay},
% \name considers the 5G and beyond protocols, but 
% only given minimal modifications. This protocol suite is
% We also implement \name using COTS
implemented upon Sivers IMA EVK06003~\cite{sivers} that operates at 60\!~GHz band and is equipped with 16-element phased arrays. 
Our major contributions are summarized as follows:
%
% with a mmWave self-interference isolation approach to extract the weak reflected signals for sensing purpose. 
%It is implemented using the off-the-shelf Sivers IMA EVK06003 platform, equipped with a 60 GHz 40-element linear phased array transmitting IEEE 802.11ad packets. Our proposed radar processing algorithms leverage channel estimation and analog / digital cancellation techniques used in a conventional IEEE 802.11ad receiver with minimal modifications. 

% However, there are several practical challenges when we want to design such a system. 1) Self-interference. In order to implement the human sensing while transmitting the data traffic, we need to extract the reflected signals from the objects. This is challenge because the self-interference will always exists. 2) Target moving. mmWave network is utilizing a directional narrow beam (so-call pencil beam) to shield a mmWave device from interference outside. It will be difficult while steering the narrow beam and sensing a moving target.
%
% \begin{figure}[!htb]
%   \centering
%   \includegraphics[width=1.05\columnwidth]{figures/phasedarray.png}
%   \vspace{-0.3cm}
%   \caption{Phased Array and weights combination}
%   \label{fig:phasedarray}
%   \vspace{-0.5cm}
% \end{figure}
%
%\vspace{-1ex}
\vspace{-0.5ex}
\begin{itemize}
    \item To the best of our knowledge, \name is the first mmWave ISAC system with 
    % the commodity 60\!~GHz front-end 
    a full-fledged sensing capability integrated with default communication function.
    %
    % \item We systematically and theoretically model the full-fledged mmWave ISAC design problems as the optimization problem. 
    \item We propose an interference cancellation scheme driven by deep neural model to combat the wideband and non-linear Tx interference to short-range sensing. 
    \item We invent an application-aware beam scheduling algorithm for jointly optimizing sensing accuracy and communication throughput.
    \item We design a unified estimation framework to leverage the sensing diversity offered by both monostatic and multi-static modes.
    \item We evaluate our \name prototype with extensive experiments. The results confirm that \name gains full-fledged mmWave ISAC capability under realistic scenarios, such as point cloud and human tracking.
\end{itemize}
Noted that \name delivers raw channel states to enable various sensing application; 
% low-level information, in particular phase to all kinds of sensing applications, but does not aim to 
it is not meant for any specific sensing purpose.
The rest of our paper is organized as follows. Section~\ref{sec:bg} provides background and motivations for mmWave ISAC systems. Section~\ref{sec:design} presents the critical components of \name. Section~\ref{sec:impl} specifies how the prototype is implemented. Section~\ref{sec:eval} reports the evaluation results on different applications. We briefly discuss general literature respectively on mmWave platforms and ISAC in Section~\ref{sec:related}, where we also explain limitations and future directions of \name.
% is presented in Section~\ref{sec:dis}, and related works are . 
Finally, Section~\ref{sec:con} concludes our paper. 

\iffalse
\paragraph{Challenges:}
\begin{itemize}
    \item 
\end{itemize}
\fi

% \needrev{
% Gemini is not designed for a specific sensing application (e.g., vital signs monitoring), but a full-fledged mmWave sensing framework to provide low-level sensing information (e.g., phase) for all kinds of sensing applications.
% }

%\paragraph{Contribution:}
%\begin{itemize}
%    \item Custom-built platform
%    \item Hybrid optimization framework
%    \item NN based interference cancellation 
%\end{itemize}

%\vspace{-1ex}
\section{Background and Motivations} 
%\vspace{-0.5ex}
\label{sec:bg}

We start with the background on mmWave communication, and we then provide motivating examples to concretely demonstrate the challenges faced by the mmWave ISAC system design.

% Both academia and industry have given significant attention to ISAC technology due to its high spectrum efficiency and low hardware cost.
\vspace{-1ex}
\subsection{Basics of mmWave Communication} \label{ssec:background}

The following procedure summarizes default mmWave communication of IEEE 802.11ay~\cite{802_11ay}. 
\begin{itemize}
	%
        %\vspace{-0.5ex}
	\item  \textbf{Carrier Sense:}  In order to avoid collisions, mmWave devices~(including AP and UEs) configure quasi-omni-directional patterns to their Rx phased arrays, and perform  listen-before-talk principle.
        %\vspace{-0.5ex}
	%
	\item \textbf{Sector Sweep}:  % The IEEE 802.11ay~\cite{802_11ay} requires finding 
 This phase determines the alignment direction of the main beams between AP and UE. 
 % in the sector-level sweep phase. 
 If channel is idle, AP broadcasts \textit{sector sweep} (SSW) frames with different narrow beams to scan all bearings, and 
 % receives 
 UEs respond with
 % corresponding 
 signal strength indicators. 
 % Hereby, the best direction between AP and UE is figured out. 
	
%	they are required by the standard to find the alignment direction of the main beams between AP and UE in the sector-level sweep phase, with each frame transmitted towards a unique direction
	
%	SSWs to all UEs, and receives their corresponding signal strength indicators. 
	%
	\item \textbf{MIMO Setup \& Training}: The AP selects UEs intended for transmissions, and sends special frames to notify them. This is followed by the AP transmitting training~(TRN) sequences to these UEs for channel state information~(CSI) estimation, and the UEs report results back to the AP.
 % \newrev{(see Appendix~\ref{append:a} for more details).} 
        %\vspace{-0.5ex}
	%
	\item \textbf{Beamforming}: The optimal MIMO configurations are set by the AP, and beamforming is performed. 
        %\vspace{-0.5ex}
	%
        %\vspace{-0.5ex}
\end{itemize}
The implications of the above procedure are twofold: 
% brings \textit{two insights} to our \name design that due to the carrier sense, 
i) only one mmWave device is allowed to transmit due to carrier sensing and later responses from an UE, and ii) the bearings of all UEs\footnote{In fact, the ranges between the AP and all UEs can be also obtained via the two-way time of arrival ranging method~\cite{ScalingmmWave_IEEE}.} are obtained in the sector sweep phase. 
Therefore, it is reasonable to consider that all UE locations are known to the AP; this applies to all subjects too (see Section~\ref{sssec:prob}).

\vspace{-1.5ex}
\subsection{Tx Interference on Short-range Sensing} \label{ssec:nfs}
\vspace{-.5ex}
%
% In an ISAC system, to enable monostatic sensing in a single mmWave transceiver is very challenge, since the Tx interference impacts on the Rx, leading to  the non-linear impaired Rx signals~\cite{}. 
%
% In an ISAC system, when attempting to transmit and receive using a single mmWave transceiver, each Tx phased array causes Tx interference that impacts the entire Rx phased array, leading to \rev{ the non-linear impaired Rx signals~\cite{}. } 
To realize the radar-like monostatic sensing function, Rx needs to be co-located with Tx to receive the reflected signals incurred by transmissions~\cite{TImmWave}, but this arrangement naturally leads to \textit{Tx interference} that potentially overpowers the reflected sensing signals. For long-range monostatic sensing envisioned for 5/6G base stations~\cite{guan-TMTT2021} where subjects are hundreds or even thousands of meters away from the signal source, one may leverage the fine-grained range bins~\cite{TImmWave} offered by the wide bandwidth of mmWave to handle Tx interference. 
%Unfortunately, short-range monostatic sensing commonly used for indoor scenarios cannot enjoy this luxury, since the strong Tx interference can ``spill out'' across several bins and hence annihilate the sensing signals. 
Unfortunately, short-range monostatic sensing, often used indoors, is challenged by strong transmission interference, which can spread across several bins and overwhelm the sensing signals.

Though the same challenge also exists for enabling ISAC upon sub-10\!~GHz IoT devices~\cite{isacot}, the Tx interference of mmWave is more complicated, due to the much higher carrier frequency and wider bandwidth. Such interference is quite different in nature too, as existing mmWave communication platforms (e.g.,~\cite{mmFLEX-MobiSys20, MIMORPH-MobiSys21}) often have separated Tx and Rx chains with their respective phased arrays, the Tx interference takes place in a \textit{cross-chain} manner, instead of the intra-chain style for IoT devices~\cite{isacot}.
%and then, even each equipped with its own phased array for performing analog beamforming, the short-range Tx interference affects the Rx chains while Tx chains transmitting. 
To better understand how Tx interference affects the short-range monostatic sensing performance, we take the ``push-pull'' hand gesture as an example, and measure the signal magnitude variations caused by it, using \name whose implementation will be introduced in Section~\ref{sec:impl}.

% The existing ISAC systems~\cite{SPARCS-IPSN22,guan-TMTT2021} do not solve the issues of TX self-interference cancellation in monostatic manner. SPARCS~\cite{SPARCS-IPSN22} is designed to work in a bistatic configuration. On the other hand, the work presented in paper~\cite{guan-TMTT2021} involves combining two separate mmWave front end devices (a transmitter and a receiver) to form a monostatic system, thereby circumventing the issue of self-interference cancellation.

% There are four main challenges that need to be addressed for successful Tx interference cancellation. When the self-interference channel is short-range dominant, the estimation of the channel and its estimation frequency are not influenced by movements in the far-field environment. 
%
% The Tx interference in far-field can be cancelled easily using transmitting delay.

% We leverage the ``push-pull'' hand gesture as an example to illustrate how Tx interference degrades its estimation performance. 
Our experiments follow 802.11ay standard~\cite{802_11ay}:
OFDM is adopted for transmission and the CSI is obtained from the 
%LTS~(long training sequence) in preamble. 
TRN field of a packet.
% \rev{TRN (training) field appended to the frame.}
For sensing purpose, we apply the inverse FFT to each CSI to obtain the CIR~(channel impulse response) as the \textit{fast-time} dimension, and combine multiple CIRs as the \textit{slow-time} dimension. For the sake of clarity, we specifically differentiate between \textit{Tx-beamforming} 
% that drives the phased array in an analog manner 
and \textit{Rx-beamforming}, where only the respective phased arrays are leveraged to achieve the ``focusing'' effect in this experiment.
% according to that processes the received signals in a digital manner.
% to represent Tx and Rx beamforming for the phased arrays of Tx and Rx chains, respectively. 

\begin{figure}[t]
\vspace{-1.5ex}
\setlength\abovecaptionskip{3pt}
\centering
\subfigure[Without proper beamforming.]{
    \includegraphics[width=0.228\textwidth]{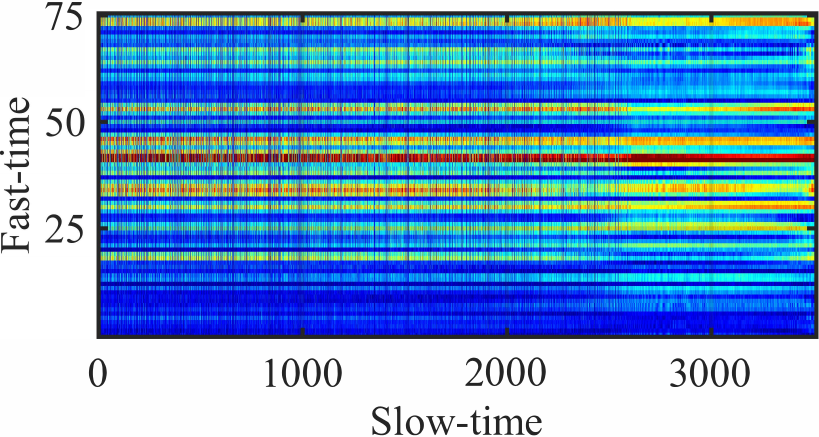}
    \label{fig:txinterfereCIR}
}
% \hspace{.1cm}
\subfigure[With proper beamforming.]{
    \includegraphics[width=0.223\textwidth]{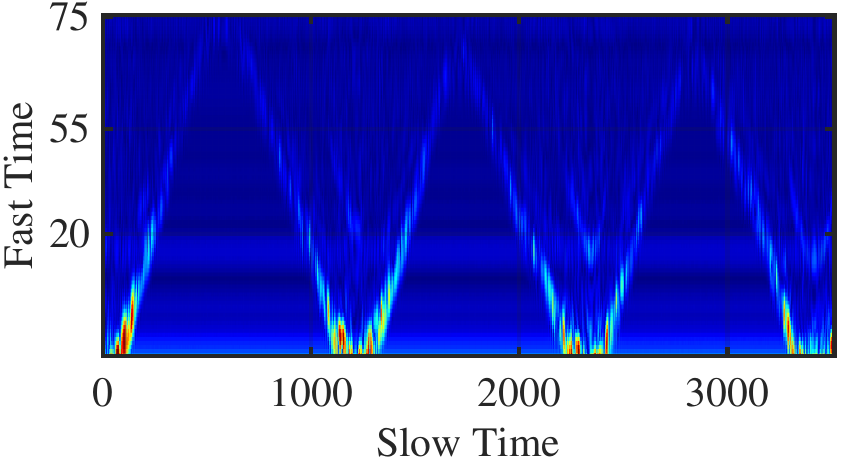}
    \label{fig:wavehandCIR}
}
    \caption{{Sensing results, in the form of CIR matrices, of push-pull hand gesture, w/ and w/o beamforming.}}
    \label{fig:TXInterfere}
    \vspace{-1.5ex}
\end{figure}

We first control the main beams of Tx-beamforming and Rx-beamforming at $0^{\circ}$ orientation to point to the hand, and show the heatmap of CIR matrix in Figure~\ref{fig:txinterfereCIR}. As expected, the heatmap exhibits random patterns that totally annihilate any features from the push-pull hand gesture, clearly demonstrating the damaging effect of the strong Tx interference to saturate the Rx chain.
% in such scenario makes the Rx chain saturated to obtain little useful sensing information. 
%
We then use 
% adjust different phase shifter patterns for beamforming and combining in the 
Tx and Rx beamforming 
%phased arrays, separately 
to cancel the Tx interference as much as possible. As shown in Figure~\ref{fig:wavehandCIR}, although certain beamforming patterns may help ``single out'' the features of the push-pull hand gesture,
%
%features using suitable beamforming and combining patterns demonstrated in the . Despite utilizing beamforming strategies to decrease the impact of Tx interference, 
minor residual interference still persists to affect the CIR matrix used for sensing. 
% This is evident in the form of ``bright lines'' in the heatmap representation of the CIR data. Apparently, it is urge to design a Tx interference cancellation solution for enabling short-range monostatic sensing in mmWave ISAC systems. 
More importantly, as beamforming is commonly used by mmWave communication systems to enhance channel quality, borrowing the same technique for sensing purpose may cause conflict between these two objectively; which leads to our next challenge.

\vspace{-1ex}
\subsection{Beam Scheduling Matters} \label{ssec:beam_schedule}
%\vspace{-0.5ex}
%
% In this section, we try to illustrate why the new beam scheduling strategy is necessary. 
Recalling that most of the current mmWave research proposals treat communication and sensing independently: they either focus on improving network throughput~\cite{mmV2X-MobiCom20,openmili-MobiCom16,nullifi-NSDI21}, or dedicate the mmWave devices to serve bistatic sensing~\cite{SPARCS-IPSN22}. Only one recent proposal~\cite{SideLobe-UbiComp23} considers simultaneous communication and sensing, but the sensing is only piggybacked on communications by leveraging the side lobes to conduct low-effective sensing:
%enable sensing but still leaves the main lobe to keep communicating, but 
it trades latency for temporal diversity so as to enhance sensing quality, at the cost of handling only static sensing subjects. 
% \rev{More importantly, all those works still inherit one-dimensional scheduling~(a.k.a. bearing scheduling) from mmWave communication for beam patterns.}
% needs to spend a lot of time to obtain good sensing information from the side lobes, and fails to sense the high temporal resolution application. 
% Moreover, in practice, the Rx mmWave devices may change their locations, and the main lobe beam tracking will be started. That results in side lobes misalignment and the algorithm of~\cite{SideLobe-UbiComp23} failure. 
For full-fledged ISAC mmWave systems, we need to balance the need from three functions, namely pure communication, pure sensing, and simultaneous communication and sensing. 
% \rev{As in Section~\ref{ssec:background}, while locations of all UEs and subjects are known, ISAC mmWave systems upgrade one-dimensional scheduling to two-dimensional scheduling~(i.e., bearings and ranges).}
% under both monostatic and bistatic modes is needed. 
In the following, we conduct two experiments using the same setting, as in Section~\ref{ssec:nfs}
%following by IEEE 802.11ad in both monostatic and bistatic modes and also use push-pull hand gesture sensing as an example 
to demonstrate the inherent conflict among these functions.
% problems in the current beam scheduling strategy. 

% Even we assume the Tx interference for short-range sensing is canceled totally, a fair scheduling algorithm that balances the three functions of pure communication, pure sensing, as well as simultaneous communication and sensing under both monostatic and bistatic modes is needed. For the conventional mmWave communication architecture, all beam scheduling strategy follows the standard protocol, such as 802.11ad. The AP starts from the beam alignment via the beamforming training procedure and then transmits data to the UE. However, in mmWave ISAC systems, such state-off-the-art beam scheduling strategy does not consider any ISAC requirement. Therefore, we conduct two experiments following by 802.11ad in both monostatic and bistatic modes and also use push-pull hand gesture sensing as an example to show the problem in current beam scheduling strategy. 

\begin{figure}[t]
\vspace{-.5ex}
\setlength\abovecaptionskip{0pt}
\centering
\subfigure[Throughput (monostatic).]{
    \includegraphics[width=0.225\textwidth]{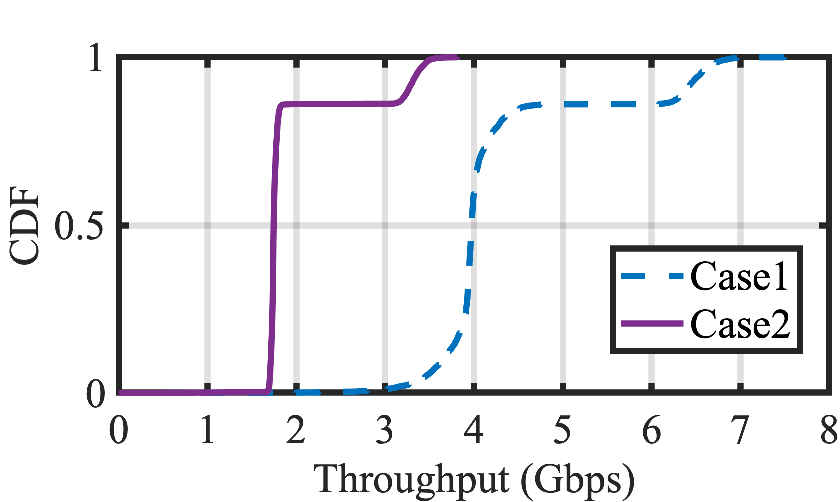}
    \label{fig:MonostaticCDF}
}
% \hspace{.1cm}
\subfigure[CIR matrix (monostatic).]{
    \includegraphics[width=0.225\textwidth]{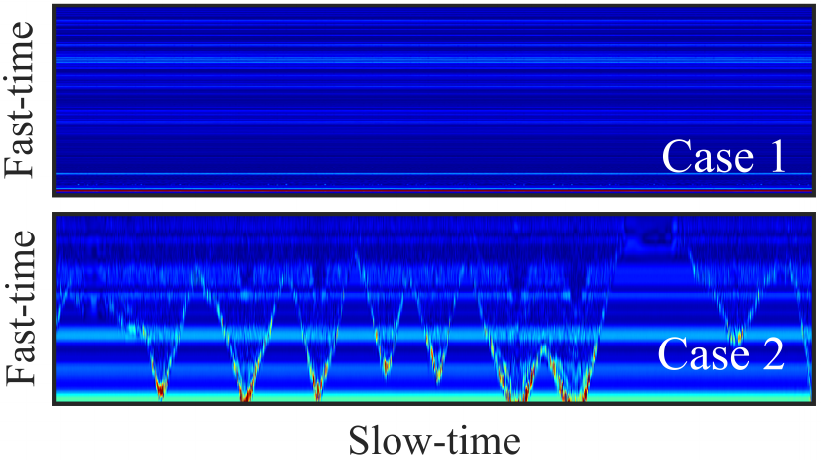}
    \label{fig:MonostaticHeatmap}
}
\subfigure[Throughput (bistatic).]{
    \includegraphics[width=0.225\textwidth]{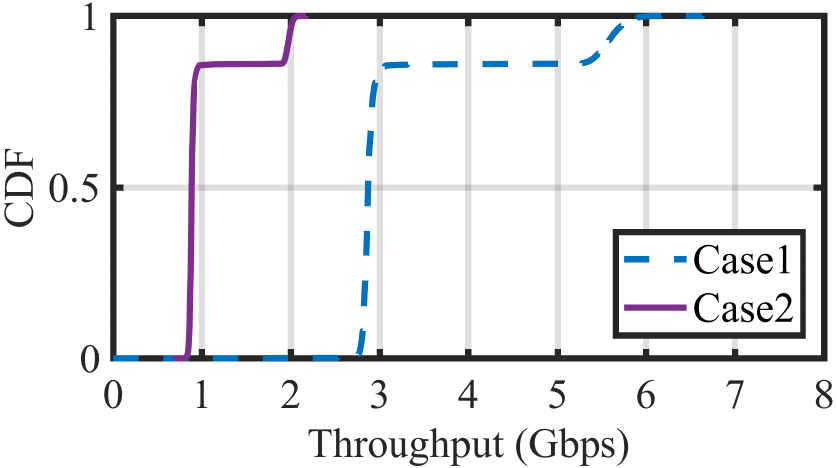}
    \label{fig:BistaticCDF}
}
% \hspace{.1cm}
\subfigure[CIR matrix (bistatic).]{
    \includegraphics[width=0.225\textwidth]{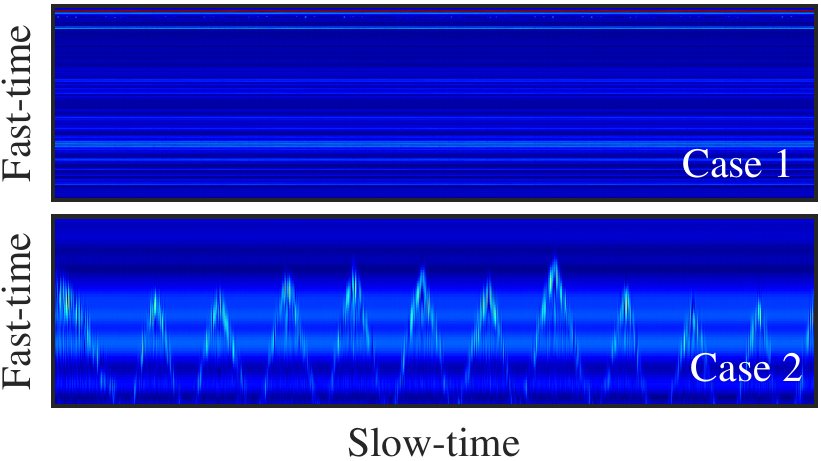}
    \label{fig:BistaticHeatmap}
}
    \caption{The communication throughput and sensing CIR heatmaps in monostatic and bistatic modes.}
    \label{fig:beam_schedul_need}
    \vspace{-2ex}
\end{figure}

% We arrange an AP, a subject, and a UE on the vertices of an isosceles obtuse triangle, with the AP and UE forming the base. 
Given a setup consisting of an AP, a subject, and a UE arranged in an isosceles triangle configuration, the performance evaluation is carried out in two distinct cases: \textsf{Case~1} chooses the best alignment between the main beams of the AP and UE for communications, while \textsf{Case~2} has the AP and UE beamforming towards the subject for sensing hand gesture (only AP Tx and Rx beamforming is needed for monostatic sensing). The performance of these cases are depicted in Figure~\ref{fig:beam_schedul_need} for both monostatic and bistatic modes. 
It is evident that \textsf{Case~1} gains much higher throughput than \textsf{Case~2} in both sensing modes, but it is the reverse situation for hand gesture sensing. 
Also, Figure~\ref{fig:BistaticHeatmap} serves as a counterexample for the effectiveness of the sidelobe sensing~\cite{SideLobe-UbiComp23}.
Apparently, the distinct beam patterns and the one-dimensional scheduling~(only in bearings) are the key factor for the performance in both cases,
% but the current communication beam scheduling strategies only can support one of communication and sensing function.
%
% since for Case~1, the main lobes of Tx and Rx beamforming are aligned to offer better communication performance, and the side lobes cannot catch enough dynamic reflections caused by the hand gesture in the both modes. On the contrary, Case~2 makes the beamforming and combines direct to the subject with hand gesture, resulting in the opposite results from Case~1. Because the main lobes from the AP and the UE is impacted heavily by the subject, leading to lower throughput for Case~2. 
% In a nutshell, the conventional communication beam scheduling cannot satisfy the requirement of ISAC. The mmWave ISAC systems need a new beam scheduling algorithm for better communication and sensing performance. 
% In a nutshell, The mmWave ISAC systems 
indicating the need for a new beam scheduling algorithm to serve the best interest of both communication and sensing under both sensing modes. 
% \rev{More importantly,  mmWave communication only consider one-dimensional scheduling~(a.k.a. bearing scheduling) for beam patterns~(see Section~\ref{ssec:background}), but .  }
%
%

\setcounter{figure}{3}
\begin{figure}[b]
\vspace{-3ex}
\setlength\abovecaptionskip{0pt}
\centering
% \hspace{.1cm}
\subfigure[Reflection power.]{
    \includegraphics[width=0.225\textwidth]{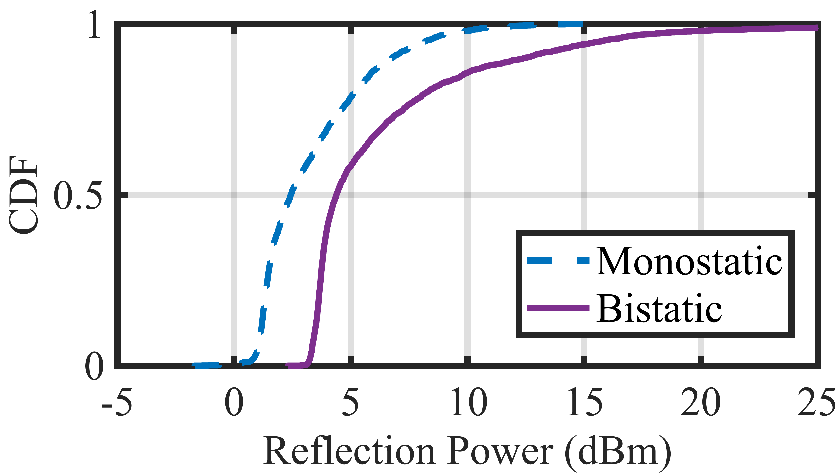}
    \label{fig:power}
}
\subfigure[S-SNR.]{
    \includegraphics[width=0.225\textwidth]{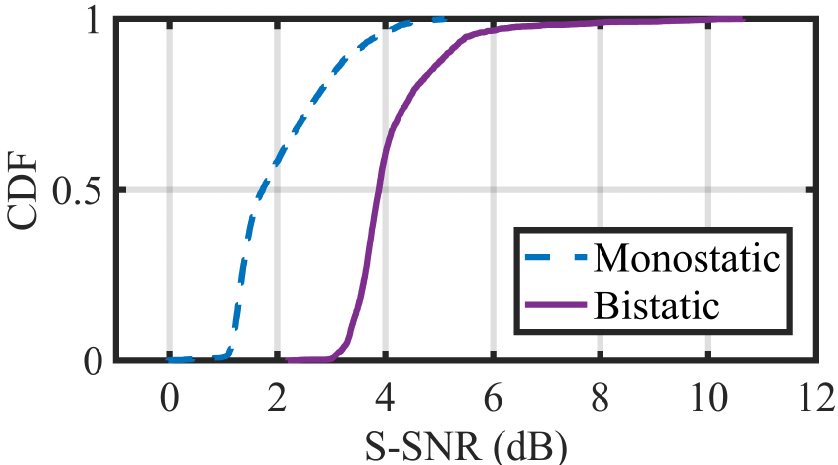}
    \label{fig:ssnr}
}
    \caption{The reflection power and S-SNR in monostatic and bistatic modes, respectively.}
    \label{fig:hybirdstatic}
    \vspace{-.5ex}
\end{figure}

\setcounter{figure}{4}
\begin{figure*}[t]
\setlength\abovecaptionskip{0pt}
  \centering
  \includegraphics[width=0.98\textwidth]{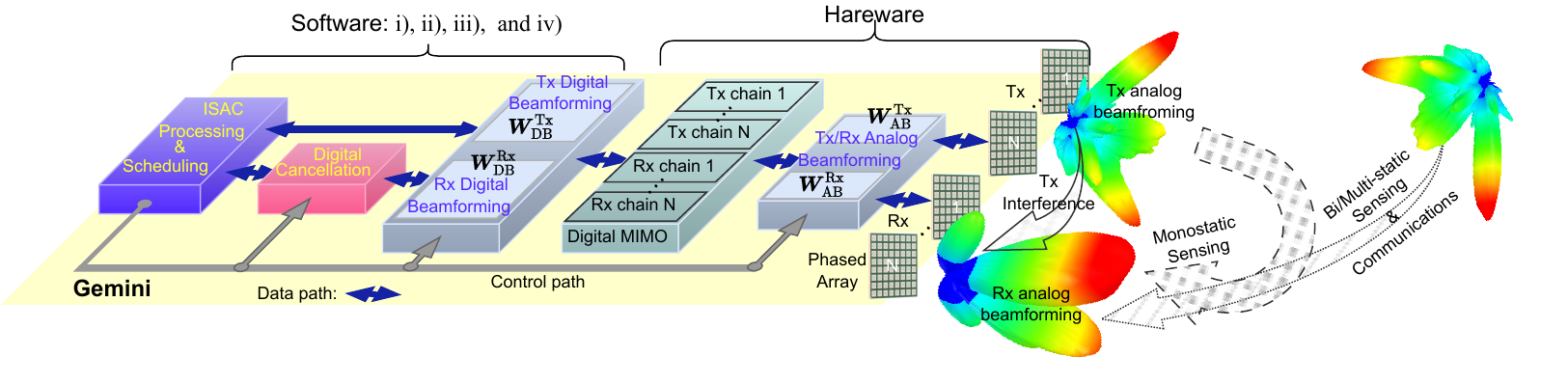}
  \caption{Architecture of \name with four software components and one hardware platform.}
  \label{fig:sys}
  \vspace{-.5ex}
\end{figure*}

%\vspace{-1ex}
\subsection{Complementary Sensing Modes} \label{ssec:com_sen_mod}
%\vspace{-.5ex}
%
% As previously noted in Section~\ref{sec:intro}, 
Whereas the monostatic mode has many advantages (mostly in terms of the synchrony between Tx and Rx)~\cite{isacot}, there exist certain situations where the bistatic mode may provide complementary sensing information: this is the \textit{diversity gain} achievable due to the different ``viewpoints''.
% In real-world mmWave ISAC scenarios, both monostatic and bistatic modes coexist, and each has its own advantages and disadvantages. 
%It's possible that a single mode may not perform optimally under certain circumstances. In this paper, we define \textit{sensing diversity} to represent more than two statistically independent complementary sensing information for concerning the same subject. In the following, we use an example to illustrate the importance of sensing diversity.
Given the same experimental setting as Section~\ref{ssec:beam_schedule}, we measure the reflection power and the signal-to-noise ratio for sensing~(S-SNR), computed similarly to that for communication) in a setup where the subject stands sideways with the shoulder facing the AP to minimize the impact of the hand gesture to the monostatic sensing signals (hence rendering it disadvantageous);
% the reflective surface area.
% During the evaluation of the bistatic mode, the experiment was conducted in a configuration where the AP, UE, and the participant were placed in an isosceles triangulation setup. This setup allowed us to evaluate the performance of the bistatic mode in a controlled environment. 
the measurements 
% We measure the reflection power and the SSNR of the push-pull hand gesture 
for both monostatic and bistatic modes are shown in Figure~\ref{fig:hybirdstatic}. 
Clearly, the bistatic mode outperforms the monostatic mode in both reflection power and SNR, with 3~\!dBm gain in average power (nearly doubling that of monostatic mode's 3.3~dBm power), as shown in Figure~\ref{fig:power}, and more than 2~\!dB gain in average S-SNR upon that of the monostatic mode's 2.05~\!dB, as shown in Figure~\ref{fig:ssnr}.

These results can be explained by the relation between the motion direction of hand and the direction of signal reflections. As motion sensing with mmWave signals relies on the signal magnitude variations caused by motion, the better the motion direction is aligned with that of reflection, the stronger the reflection variations are. In reality, the chances for an arbitrary motion direction to be aligned with either monostatic or bistatic sensing mode is surely higher than with only one of them, because combining distinct reflection directions could improve the signal diversity.
%
% reason is that in this case, for the monostatic mode, the angle between incidence and reflection signals is too narrow to project a good RCS~(radar cross-section)~\cite{principle_of_modern_radar}. Still, the bistatic mode has the larger angle of those leading to better RCS. Larger RCS means the subject can be more easily detected. 
%
% The monostatic mode has a lower SSNR than the bistatic mode, as depicted in Figure~\ref{fig:ssnr}. Additionally, the bistatic mode has a lower noise power of 0.79 dBm, while the monostatic mode has a relatively higher noise level of up to 4.31 dBm due to interference. The average reflection power in the bistatic mode (6.3dBm) is significantly higher than that in the monostatic mode (3.3dBm), nearly twice as much.
% Although the monostatic mode provides an advantage in accurately estimating the location of a subject's hand motion through real-time flight (ToF) ~\cite{principle_of_modern_radar}, measurement, it has limitations, and the bistatic mode may be more appropriate in certain situations. However, the bistatic mode does introduce additional overhead compared to the monostatic mode for hand motion estimation~\cite{SPARCS-IPSN22}. 
Since an ISAC system should definitely leverage this sensing diversity
% from both monostatic and bistatic modes 
to offer effective sensing capabilities,
% for the subjects. Although our naive example demonstrates that sensing diversity 
% improve that performance, 
it is imperative to have a unified framework for merging monostatic and multi-static sensing modes 
% to tackle the sensing diversity information 
in a constructive way.

\vspace{-1ex}
\section{\name: {\small mm}Wave ISAC Design} 
\vspace{-.5ex}
\label{sec:design}

%=============== Outlines ================
%1.hardware platform
%2.control path
%3.data path

Motivated by the observations made in Section~\ref{sec:bg}, our \name design comprises five key components: i) a sensing-aware channel probing scheme, ii) a two-stage Tx interference cancellation, iii) a holistic beamforming and scheduling mechanism for both sensing and communications, iv) an algorithm to exploit the diversity in sensing modes, 
%v) an ISAC protocol compatible with mmWave Wi-Fi, 
and v) a hardware platform to support previous components. Given the overall construction of \name shown in Figure~\ref{fig:sys}, the first four components are held in the three leftmost blocks, and they control the remaining (mostly hardware) blocks to perform beamforming and cancel interference, aiming to satisfy the diversified ISAC requirements. 
%
% \name is also co-designed tightly by hardware and software. From the right side to the left side, The phased arrays and their Tx/Rx chains are used to construct the hybrid Tx/Rx beamforming. The digital cancellation is leveraged to reduce the residue Tx interference after hybrid beamforming. The ISAC process and scheduling are the controller of \name that control all components and tackle the data.
%
In the following, we introduce the first four components respectively but postpone the platform implementation details to Section~\ref{sec:impl}.

% the rose red and blue blocks are the hardware and the software parts, respectively. It is noted that both hardware and software are co-designed tightly. The phased arrays and the Tx/Rx chains are used to construct the hybrid Tx/Rx beamforming. \needrev{The analog and digital cancellation are controlled by the ISAC process and schedule that are elaborated in Section~\ref{ssec:beam_design_sche}. The ISAC process and scheduling also runs the unified sensing framework based on sensing diversity to improve the sensing performance introduced in Section~\ref{ssec:sens_dive}. }

% Since \name is designed and developed as a general-purpose system for deploying mmWave ISAC devices that is flexible, cost-effective, and efficient, its main components need to easily adapt to diverse requirements. The current mmWave platforms are bulky and complicated, so \name's system will be configurable and able to reuse hardware resources in order to overcome these challenges in ISAC. The goal is to create a system that can be easily deployed without incurring additional costs from new hardware. In this section, we describe how we design the \name ISAC system that is composed of a series modules. 

% 1. channel model
% 2. modeling the problem : 1. Tx interference, 2. communication, 3. sensing together
% 3. introduce the multiple nodes sensing cooperation. sensing diversity
% 4. conclusion 
\vspace{-1ex}
\subsection{Channel Modeling and Probing} \label{ssc:model_prob}
%\vspace{-0.5ex}

In this section, we design a probing scheme to obtain mmWave MIMO channel states for both sensing and communication, based on a properly defined channel model.

\vspace{-1ex}
\subsubsection{Modeling mmWave MIMO Channels} \label{sssc:model_chan}
\vspace{-0.5ex}
%
% Before we design a channel probing method for mmWave ISAC channels estimation, we need to dive into the mmWave MIMO reception model to understand them better. 
Compared with conventional (communication) channel, the mmWave MIMO channels for ISAC can be far more complicated, as they include at least four components: Tx interference, monostatic sensing (reflection), communication, and multi-static sensing (reflection). 
% \needrev{Although the work~\cite{adib2013see} also leverages the beamforming to null Tx interference for monostatic sensing at 2.4~\!GHz, it does not consider communication and multi-static sensing.  }
%Considering the standard protocol, such as IEEE 802.11ay compatibility, there is still half-duplex communication mode. 
%
In a typical indoor mmWave MIMO setting, the AP has $N$ Tx/Rx chains, each of them equipped with an $M$-elements phased array. 
% For simplicity, we consider the sum of the number of UEs and the subjects are equal to $N = N_{\mathrm{U}} + N_{\mathrm{S}}$, and each UE only has one Tx/Rx chain with a phased array. 
% In this paper, we utilize the word ``users'' to represent both UEs and subjects. 
For ISAC-oriented temporal scheduling, the AP selects $N_{\mathrm{U}}$ UEs and $N_{\mathrm{S}}$ subjects (where $N = N_{\mathrm{U}} + N_{\mathrm{S}}$) from all UEs and subjects to serve at each time slot. 
% The AP performs MU-MIMO to the UEs for the IEEE 802.11ay, but the only one UE is allowed to communicate with the AP in the uplink case, due to the CSMA/CA mechanism adopted by the IEEE 802.11ay. 
Since the MU-MIMO via hybrid beamforming is adopted by IEEE 802.11ay~\cite{802_11ay}, we consider the AP performing the MU-MIMO to the UEs. 
We let $\boldsymbol{s}(t) = [ s_1(t), \cdots, s_N(t) ]$, $\boldsymbol{y}^{\mathrm{u}}(t) = [y^{\mathrm{u}}_1(t), \cdots, y^{\mathrm{u}}_{N_{\mathrm{U}}}(t) ]$, and  $\boldsymbol{y}^{\mathrm{s}}(t) = [y^{\mathrm{s}}_1(t), \cdots, y^{\mathrm{s}}_{N_{\mathrm{S}}}(t) ]$ respectively represent the Tx signals from the AP, the Rx signals at the UEs, and the monostatic sensing signals received by the AP. Note that, $\boldsymbol{y}^{\mathrm{u}}(t)$ and $\boldsymbol{y}^{\mathrm{s}}(t)$, even when happening to the same device, are temporally separated by the MAC protocol (see Section~\ref{ssec:background}), while $\boldsymbol{y}^{\mathrm{u}}(t)$ is the superposition of the multi-static sensing signals $\boldsymbol{y}^{\mathrm{us}}(t)$ and the pure communication signals $\boldsymbol{y}^{\mathrm{uc}}(t)$, i.e., $\boldsymbol{y}^{\mathrm{u}}(t) = \boldsymbol{y}^{\mathrm{us}}(t) + \boldsymbol{y}^{\mathrm{uc}}(t)$. In total, we can characterize the simultaneous ISAC (both AP and UE) Rx signals by:
\begin{align} \label{eq:rx_sigs}
   \boldsymbol{y}(t) &= [\boldsymbol{y}^{\mathrm{u}}(t), \boldsymbol{y}^{\mathrm{s}}(t)]' \cr
    &= \boldsymbol{W}_{\mathrm{DB}}^{\mathrm{Rx}} \boldsymbol{W}_{\mathrm{AB}}^{\mathrm{Rx}} \boldsymbol{H}(t) \boldsymbol{W}_{\mathrm{AB}}^{\mathrm{Tx}} \boldsymbol{W}_{\mathrm{DB}}^{\mathrm{Tx}}  \boldsymbol{s}(t) + \boldsymbol{W}_{\mathrm{DB}}^{\mathrm{Rx}} \boldsymbol{W}_{\mathrm{AB}}^{\mathrm{Rx}} \boldsymbol{n}(t),
\end{align}
where $\boldsymbol{W}_{\mathrm{DB}}^{\mathrm{Rx}}$, $ \boldsymbol{W}_{\mathrm{DB}}^{\mathrm{Tx}} $, $ \boldsymbol{W}_{\mathrm{AB}}^{\mathrm{Rx}} $, and $\boldsymbol{W}_{\mathrm{AB}}^{\mathrm{Tx}} $  are the Rx digital beamformers, the Tx digital beamformers, the Rx analog beamformers, and the Tx analog beamformers, respectively, $[\cdot]'$ denotes matrix transpose, and $\boldsymbol{n}(t)$ is the additive Gaussian noise with variance $\sigma^2$. It is worth noting that $\boldsymbol{W}_{*}^{\mathrm{Rx}}$ is of block-diagonal form, as it involves the beamformer of one device (e.g., UE) for $\boldsymbol{y}^{\mathrm{u}}(t)$ and that of another (e.g., AP) for $\boldsymbol{y}^{\mathrm{s}}(t)$.

The $\boldsymbol{H}(t)$ in Eqn.~\eqref{eq:rx_sigs} represents the mmWave MIMO channel; it involves the following major components:
\vspace{-0.5ex}
\begin{align} \label{eq:csi}
   \boldsymbol{H}(t) &= \boldsymbol{H}_{\mathrm{ti}}(t) + \boldsymbol{H}_{\mathrm{s}}(t) + \boldsymbol{H}_{\mathrm{c}}(t) + \boldsymbol{H}_{\mathrm{ms}}(t),  
\vspace{-1.5ex}
\end{align}
where the 4 terms on the right-hand side 
% the $\boldsymbol{H}_{\mathrm{ti}}(t) $, $\boldsymbol{H}_{\mathrm{s}}(t) $, $\boldsymbol{H}_{\mathrm{uc}}(t) $ and $\boldsymbol{H}_{\mathrm{us}}(t)$ 
denote the Tx interference channels, monostatic sensing channels, communication channels, and multi-static sensing channels, respectively. According to Section~\ref{ssec:nfs}, $\boldsymbol{H}_{\mathrm{ti}}(t)$ may strongly affect $\boldsymbol{H}_{\mathrm{s}}(t)$, but the impact from $\boldsymbol{H}_{\mathrm{c}}(t)$ to $\boldsymbol{H}_{\mathrm{ms}}(t)$ can be readily handled~\cite{mmtrack-INFOCOM20}.
% on the communication channels and the multi-static sensing channels due to the larger distance separation. 
Moreover, Eqn.~\eqref{eq:rx_sigs} confirms what we have observed in Section~\ref{ssec:beam_schedule}: the hybrid beamformers can be scheduled to largely reshape the channels so as to affect performance in both communication and sensing.
Consequently, though the two groups of channels, i.e., $\{\boldsymbol{H}_{\mathrm{ti}}(t), \boldsymbol{H}_{\mathrm{s}}(t)\}$ vs $\{\boldsymbol{H}_{\mathrm{c}}(t), \boldsymbol{H}_{\mathrm{ms}}(t)\}$, are seemingly independent of each other as they are physically or temporally separated, they become correlated as they may share the same Tx beamformers $\boldsymbol{W}_{*}^{\mathrm{Tx}}$. 
%
% due to the large number of antennas in the phased array~(e.g. $32\times32$-antennas phased arrays for each Tx/Rx chain in our platform), we are hard to figure out the optimal hybrid beamformers for all types channels in Eqn.~\eqref{eq:csi} via a one-shot efficient algorithm.
In a nutshell, the goal of \name is to \textit{optimize the hybrid Tx/Rx beamformers on all these channels in $\boldsymbol{H}(t)$ for balancing the performance between sensing and communication}. 
%
% Therefore, in the following section, we will divide the above mmWave MIMO channels Eqn.~\eqref{eq:csi} into two parts and propose a probing method to estimate them for empirically tackling the corresponding problems step by step
%
We take a divide-and-conquer method to approach this optimization, by eliminating $\boldsymbol{H}_{\mathrm{ti}}(t)$ in Section~\ref{ssec:mono_sens} and jointly scheduling $\{\boldsymbol{H}_{\mathrm{s}}(t), \boldsymbol{H}_{\mathrm{c}}(t), \boldsymbol{H}_{\mathrm{ms}}(t)\}$ in Section~\ref{ssec:sche_isac}.
However, before executing this plan,
% diving into the beamformer optimization, 
a scheme to probe the channel states needs to be in place.
%
% is extraordinary challenge via a one-shot optimization formulation and solution. 

\vspace{-0.5ex}
\subsubsection{Probing mmWave ISAC Channels} \label{sssec:prob}
\vspace{-0.5ex}
%
%0. only consider the two channels Hsi and Hs
%1. measurement for the Tx interference : there are three regions, heavy interference without any sensing information; Tx/Rx beamforming cancellation with sensing information; non-Tx interference but no sensing information
%2. solve the optimization problems for hybrid beamforming
%3. flow-based dnn interference denoise
%By default, mmWave communication systems should have been equipped with a probing scheme for acquiring $\boldsymbol{H}_{\mathrm{c}}$ as required by IEEE 802.11ay~\cite{802_11ay}. 
Recall that a probing scheme is employed for mmWave communications to acquire $\boldsymbol{H}_{\mathrm{c}}$ in Section~\ref{ssec:background} under the sector sweep phase, which also yields $\boldsymbol{H}_{\mathrm{ms}}(t)$ and the \textit{bearing}s of subjects via static background removal~\cite{adib2015multi} and signal 
% a well-known constant false alarm rate 
detection~\cite{richards2014fundamentals}.
Since 
% there is no such a scheme for acquiring 
$\{\boldsymbol{H}_{\mathrm{ti}}(t), \boldsymbol{H}_{\mathrm{s}}(t)\}$ cannot be directly acquired by this scheme, we piggyback an additional probing scheme on the existing one by emulating its behavior.
%
%According to Eqn.~\eqref{eq:csi}, we realize that the Tx interference and monostatic channels happen in the AP; the communication and multi-static sensing channels exist in the UEs. For the latter one, the UEs can estimate the channels followed by probing mechanism of the IEEE 802.11ay, and transmit them via a feedback frame to the AP, but the former one does not has the mechanism support so far. 
Consequently, we mainly focus on this added scheme 
% and hence we slightly abuse terminology by having $\boldsymbol{H}(t) = \boldsymbol{H}_{\mathrm{ti}}(t) + \boldsymbol{H}_{\mathrm{s}}(t)$; this scheme 
aiming to estimate the Tx interference channels $\boldsymbol{H}_{\mathrm{ti}}(t)$ for the follow-up cancellation that obtains monostatic sensing channel $\boldsymbol{H}_{\mathrm{s}}(t)$ in Section~\ref{ssec:mono_sens}. 
%
% In practice, discovering the sensing channels $\{\boldsymbol{H}_{\mathrm{s}}(t), \boldsymbol{H}_{\mathrm{ms}}(t)\}$ would not be purely ``discovered'' as $\boldsymbol{H}_{\mathrm{c}}$, but 
A byproduct of this two-round probing and the later resulting $\boldsymbol{H}_{\mathrm{s}}(t)$ is the locations of all subjects, as subject \textit{range}s can be inferred from $\boldsymbol{H}_{\mathrm{s}}(t)$'s time-of-flight.
% can be estimated via  in the sector sweep phase. Consequently, the subjects' directions are known to \name, so their channel states can be obtained by directly focusing on those directions.

% \rev{Both the bearing and range of a subject are estimated by a well-known constant false alarm rate detection algorithm~\cite{richards2014fundamentals} in the sector sweep phase~(see Section~\ref{ssec:background}).}

% \needrev{In practice, the sensing channels $\{\boldsymbol{H}_{\mathrm{s}}(t), \boldsymbol{H}_{\mathrm{ms}}(t)\}$ would not be purely ``discovered'' as $\boldsymbol{H}_{\mathrm{c}}$, because a (sensing) subject should subscribe to \name's service first. Consequently, the subject's direction is known to \name, so its channel state can be obtained by directly focusing on that direction.}

% Since the monostatic reflection signals are received by the AP itself, to cancel the Tx interference for monostatic sensing, we consider only Tx interference and monostatic sensing channels, and slightly abuse the symbol $\boldsymbol{H}(t) = \boldsymbol{H}_{\mathrm{ti}}(t) + \boldsymbol{H}_{\mathrm{s}}(t)$. Our goal is to estimate the monostatic sensing channels $\boldsymbol{H}_{\mathrm{s}}(t)$, and the Tx interference channels $ \boldsymbol{H}_{\mathrm{ti}}(t) $.
% but suppress the Tx interference channels $ \boldsymbol{H}_{\mathrm{ti}}(t) $. Naturally, The prior task before Tx interference cancellation is to catch  that channels $ \boldsymbol{H}_{\mathrm{ti}}(t)$.
%
% The probe is used to obtain the Tx interference channels for the following cancellation. 

Since the phased array equipped at Tx chain has multiple beam patterns and each of them causes different Tx interference, a probing scheme needs to collect all kinds of corresponding Tx interference channels. 
%\rev{Fortunately, this can be achieved via slightly modifying the sector sweep phase~(see Section~\ref{ssec:background}).} 
% \needrev{ Fortunately, this can be achieved via \textit{sector sweep} (SSW) frames~\cite{802_11ay}:
% they are required by the standard to find the alignment direction of the main beams between AP and UE in the sector-level sweep phase, with each frame transmitted towards a unique direction specified by $\boldsymbol{W}_{AB}^{\mathrm{Tx}}$. }
% of Tx analog beamforming, leading to different Tx interference conditions. 
%
Therefore, our added scheme involves an additional round of SSW with a substantially lower Tx power to create a ``wireless shortcut'' for probing $\boldsymbol{H}_{\mathrm{ti}}(t)$, in order to avoid the interference from surrounding environments.
% and obtain pure Tx interference
% attenuate the signals quickly while they go beyond 1\!~m.   
As shown in Figure~\ref{fig:probe}, upon transmitting SSW frames with Tx sector ID sequentially from a device~(the pink sectors), the phased arrays of its Rx chains are set to quasi-omni-directional patterns to receive SSW frames~(the blue sectors). 
% The probe is carried out at both the UE and the AP sides. The Tx (illustrated in pink color) sends SSW frames in different directions, and the Rx with its quasi-omni-directional beam pattern (represented in light blue color) will receive the probe SSW frames from the Tx, as shown in Figure~\ref{fig:probe}. 
In this way, the channel estimation field with TRN
%training sequence 
of SSW is utilized to extract the Tx interference channels $\boldsymbol{H}_{\mathrm{ti}}(t)$ via a minimum mean square-error estimation~\cite{speth1999optimum}.
\setcounter{figure}{5}
\begin{figure}[t]
    %\setlength\abovecaptionskip{8pt}
    %\vspace{-1em}
    \centering
    \includegraphics[width=1\columnwidth]{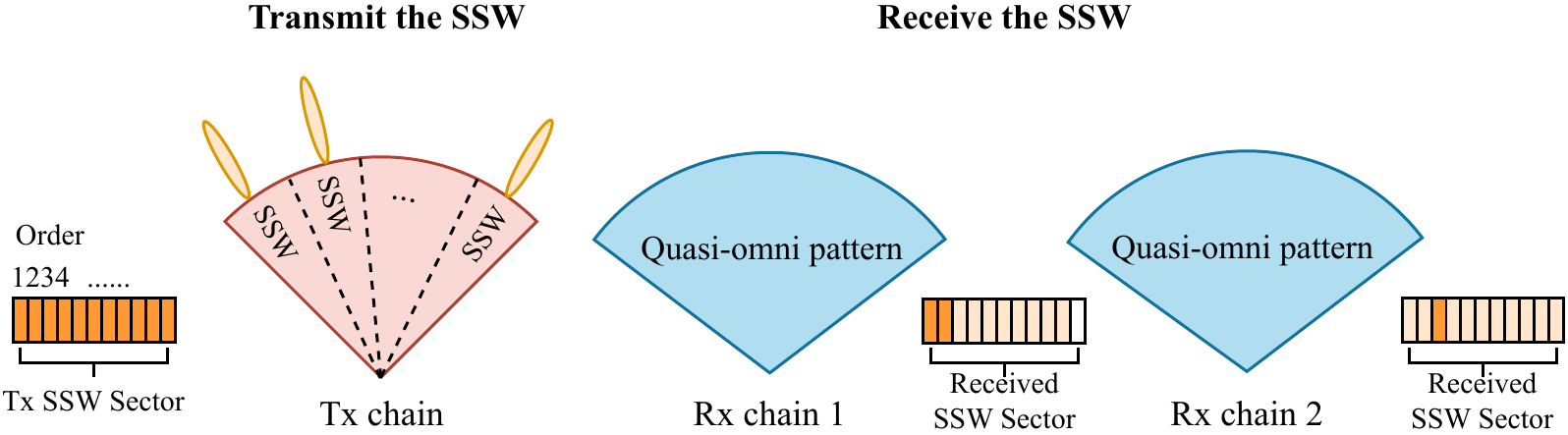}
    \caption{Tx interference probing via extended SSW: ``SSW sector'' bars show the interference strength for each 
    % channel and 
    direction.
    % with deeper orange showing more substantial interference power and lighter colors indicating weaker interference levels. 
    }
    \label{fig:probe}
    \vspace{-1ex}
\end{figure}
\begin{figure}[t]
    \vspace{-.5ex}
    \setlength\abovecaptionskip{0pt}
    \subfigure[Comunication.]{
        \includegraphics[width=0.45\linewidth]{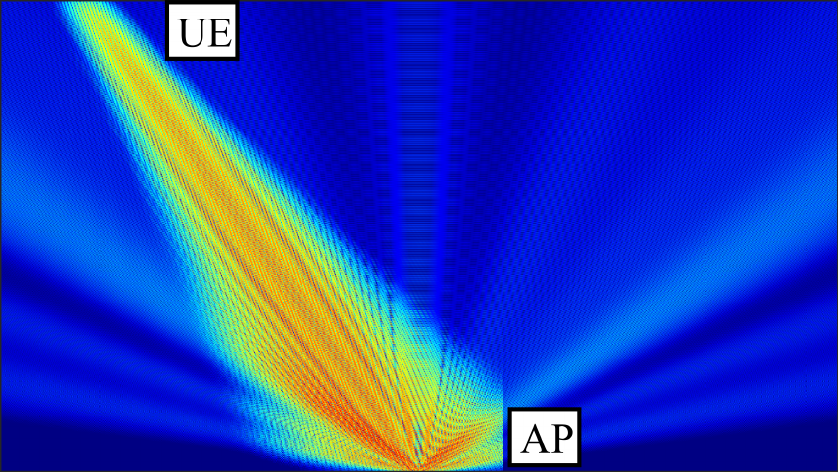}
        \label{fig:commBeam}
    }
    \subfigure[Bistatic.]{
        \includegraphics[width=0.45\linewidth]{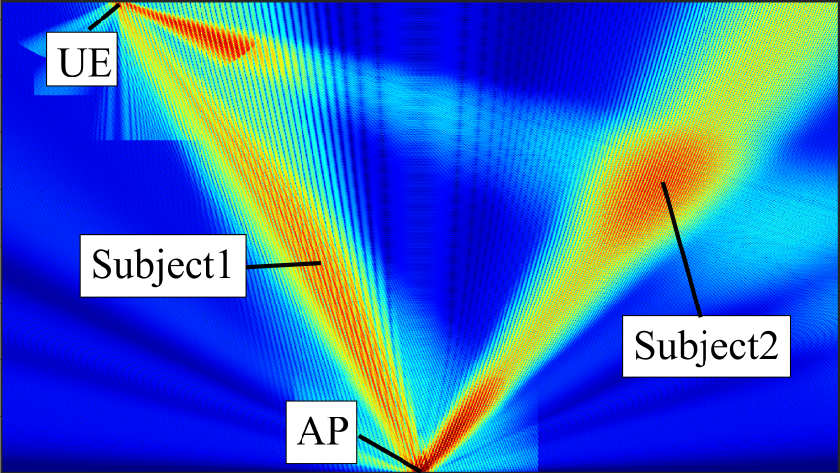}
        \label{fig:mul_probe}
    } 
    \\\vspace{-1em}
    \subfigure[Tx interference for monostatic.]{
        \includegraphics[width=0.45\linewidth]{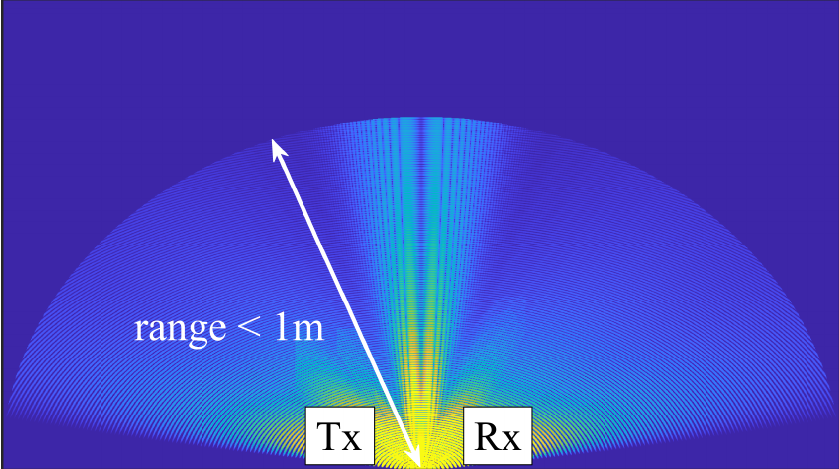}
        \label{fig:mono_probe}
    }
    \subfigure[Tx interference correlation.]{
        \includegraphics[width=0.45\linewidth]{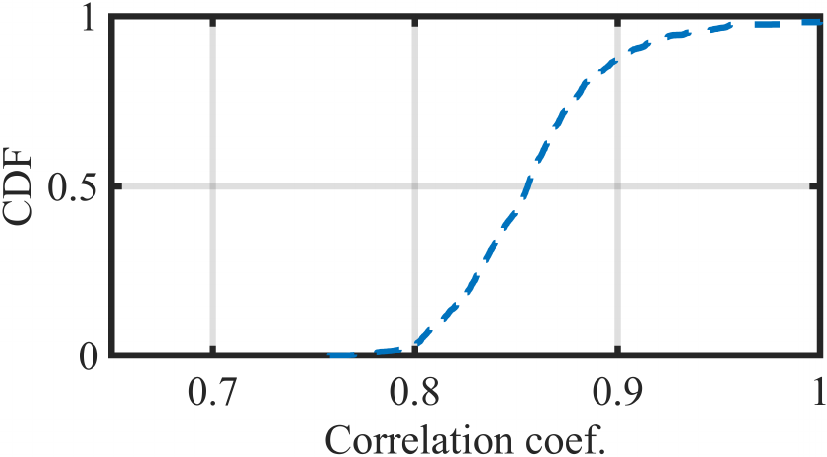}
        \label{fig:TxinterCorr}
    }  
    %\vspace{-1ex}
    \caption{Beamforming heatmaps of (a-b) default probing for communications and multi-static sensing, (c) new probing scheme for Tx interference, with (d) statistics on interference stability.} 
    \label{fig:probing4}
    \vspace{-2ex}
\end{figure}
%We perform the aforementioned two-round SSW in an example setting: a UE appears at -35$^\circ$ bearing with respect to an AP, with two subjects at respectively -25$^\circ$ and 30$^\circ$; the results are reported in Figure~\ref{fig:probing4} as heatmaps accumulating all probing SSW frames.

%
% The group of communication and bistatic sensing channels $\{\boldsymbol{H}_{\mathrm{c}}(t), \boldsymbol{H}_{\mathrm{ms}}(t)\}$ are acted on the UE, and less impacted by the $\boldsymbol{H}_{\mathrm{ti}}$, but they affect the $\boldsymbol{H}_{\mathrm{s}}(t)$ occurred at the AP. Therefore, we have the opportunity to divide the mmWave MIMO channels Eqn.~\eqref{eq:csi} into two parts as illustrated in Section~\ref{sssc:model_chan} and perform scheduling for both communication and sensing via sharing Tx beamformers. 
%
Whereas the UE channel in Figure~\ref{fig:commBeam} is discovered by the 1st round SSW, the two sensing channels in Figure~\ref{fig:mul_probe} are probed during the same SSW round. 
% according to the known directions. 
Apparently, the beamforming for UE may allow for sensing `Subject1' at the same time, yet sensing `Subject2' has to be scheduled in a different time slot. 
% the communication and bistatic sensing cases, the UE and subject are in the same direction, and the current scheduling algorithms used in communication assign different time slots to perform either communication or bistatic sensing. Those scheduling algorithms not only waste the time resource, but also are very inefficient in mmWave ISAC scenario.
%
Moreover, the 2nd round SSW obtains Tx interference states shown in Figure~\ref{fig:mono_probe}: the impact appears to be more intensive at 0$^\circ$ and is otherwise quasi-symmetric on two sides.
% It illustrates the power of Tx interference impacts heavily on the Rx chain, especially the red color area in the middle, and due to the lower Tx power, the probing scheme can obtain the nearly pure Tx interference in the short range. 
We also measure the correlation between any two Tx interference channels at different time slots, and the results in Figure~\ref{fig:TxinterCorr} demonstrate correlation coefficients mostly above 0.8. Therefore, the information provided by one round SSW can help train a cancellation process that remains valid for a long period.
% This observation indicates we can design a cancellation solution to suppress the stable Tx interference channels, and extract the monostatic sensing channels in the following.
%

\vspace{-1.2ex}
\subsection{Monostatic Sensing for mmWave ISAC }  \label{ssec:mono_sens}
\vspace{-0.5ex}
In this section, we introduce a two-stage Tx interference cancellation process to enable monostatic sensing: namely beam nulling and deep denoising.

\vspace{-0.9ex}
\subsubsection{Beam Nulling}
\vspace{-0.5ex}

We leverage the hybrid beamforming to reduce the Tx interference at the first stage. 
Since advanced phased array allows for controlling both amplitude and phase~\cite{SiversBF01_2021} via antenna weights vectors (AWV), we leverage this ability to offer efficient solutions for future developments.
% we do not consider the legacy phased array with only phase controlling.
% \needrev{Common analog beamformers $\boldsymbol{W}_{\mathrm{AB}}^{\mathrm{Tx}}$ and $\boldsymbol{W}_{\mathrm{AB}}^{\mathrm{Rx}}$ for communications control only the phase of output but not the amplitude~\cite{nullifi-NSDI21}.} 
% Therefore, in order to offer efficient solutions for future developments, we first relax this constraint to obtain analog beamformers capable of controlling both phase and amplitude. 
To achieve beam nulling for the $i$-th sector containing at least a subject, we leverage
% formulate an optimization problem using 
the hybrid beamforming to minimize Tx interference $\boldsymbol{H}^i_{\mathrm{ti}}(t)$:
%
% \begin{align} \label{eq:m-ssnr}
%    \max \quad &  \left \Vert
% \boldsymbol{W}_{\mathrm{DB}}^{\mathrm{Rx}} \boldsymbol{W}_{\mathrm{AB}}^{\mathrm{Rx}} \boldsymbol{H}_{\mathrm{s}}(t) \boldsymbol{W}_{\mathrm{AB}}^{\mathrm{Tx}} \boldsymbol{W}_{\mathrm{DB}}^{\mathrm{Tx}}  \right \Vert^2_F \\
%     \mathrm{s.t.}  \quad  & \Vert  \boldsymbol{W}_{\mathrm{DB}}^{\mathrm{Rx}}\boldsymbol{W}_{\mathrm{AB}}^{\mathrm{Rx}} \boldsymbol{H}_{\mathrm{ti}}(t) \boldsymbol{W}_{\mathrm{AB}}^{\mathrm{Tx}}\boldsymbol{W}_{\mathrm{DB}}^{\mathrm{Tx}}  \Vert^2_F = 0
% \end{align}
%
\begin{align} \label{eq:m-ssnr}
   \textstyle{\min_{\boldsymbol{W}_{\mathrm{DB}}^{\mathrm{Rx}},\boldsymbol{W}_{\mathrm{AB}}^{\mathrm{Rx}}, \boldsymbol{W}_{\mathrm{AB}}^{\mathrm{Tx}},\boldsymbol{W}_{\mathrm{DB}}^{\mathrm{Tx}}}} ~~&  \Vert  \boldsymbol{W}_{\mathrm{DB}}^{\mathrm{Rx}}\boldsymbol{W}_{\mathrm{AB}}^{\mathrm{Rx}} \boldsymbol{H}^i_{\mathrm{ti}}(t) \boldsymbol{W}_{\mathrm{AB}}^{\mathrm{Tx}}\boldsymbol{W}_{\mathrm{DB}}^{\mathrm{Tx}}  \Vert^2_F,
\end{align}
where $\Vert \cdot \Vert_F$ is the Frobenius norm.
% $\eta_{\mathrm{s}} = \left \Vert
% \boldsymbol{W}_{\mathrm{DB}}^{\mathrm{Rx}} \boldsymbol{W}_{\mathrm{AB}}^{\mathrm{Rx}} \boldsymbol{H}_{\mathrm{s}}(t) \boldsymbol{W}_{\mathrm{AB}}^{\mathrm{Tx}} \boldsymbol{W}_{\mathrm{DB}}^{\mathrm{Tx}}  \right \Vert^2_F \\$.  
Since the hybrid beamformers involve a large amount of parameters to be optimized, searching for an optimal solution directly using conventional optimization methods can be highly inefficient.
Moreover, since these parameters will also be needed for the joint scheduling in Section~\ref{ssec:sche_isac}, the degree of freedom in fine-tuning them is limited.
% each path of Tx interference channels is comprised of amplitude, time delay, angle of arrival, angle of departure, etc., the search space of this problem is too large to accommodate an efficient solution. 

\begin{figure}[t]
\vspace{-.5ex}
  \setlength\abovecaptionskip{8pt}
  \centering
  \includegraphics[width=.86\columnwidth]{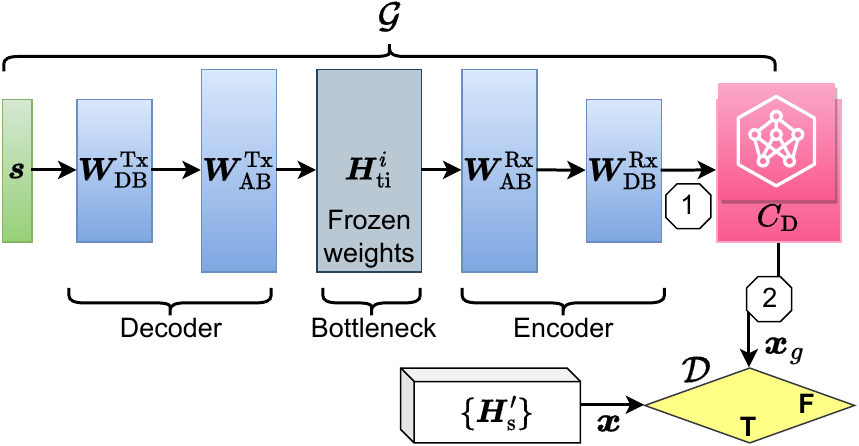}
  % \vspace{-0.3cm}
  \caption{The two-stage Tx interference cancellation via \textit{linear autodecoder} (LAE) and cGAN.}
  \label{fig:autodecoder}
  \vspace{-2em}
\end{figure}

Fortunately, we find the problem~\eqref{eq:m-ssnr} can be approached as training a neural network, given the similarity between the data pipeline (see Figure~\ref{fig:sys}) and a linear autoencoder~\cite{kunin2019loss}. We reformulate Eqn.~\eqref{eq:m-ssnr} into an inverse neural network shown in Figure~\ref{fig:autodecoder}, where the Tx interference channels and the Tx/Rx hybrid beamformers are respectively modeled as the bottleneck layer (weights frozen as the channel $\boldsymbol{H}^i_{\mathrm{ti}}(t)$ is physically determined) and decoder/encoder whose weights are beamformer parameters,
%inverse architecture of that to  in neural network and solve it efficiently. In stark contrast to the conventional autoencoder that is a downsampling and then upsampling architecture, our autodecoder reverses its architecture shown in.
while the Tx signals $\boldsymbol{s}$ become the TRN.
%training sequences. 
The objective of a hypothetical training process should result in the output $\boldsymbol{x}$ being the $\boldsymbol{H}_{\mathrm{s}}(t)$ after canceling the overwhelming $\boldsymbol{H}^i_{\mathrm{ti}}(t)$. However, 
% since we do not have
the network cannot be trained without any ground truth dataset for $\boldsymbol{H}_{\mathrm{s}}(t)$.

\vspace{-0.5ex}
\subsubsection{Deep Denoising}
\vspace{-0.5ex}
%
%For the heavy Tx interference area, specifically \textsf{Sector~1}, the hybrid beamforming cancellation is insufficient to suppress it. 
By far, we have reformed the problem to reconstructing $\boldsymbol{H}_{\mathrm{s}}(t)$ from the observation $\boldsymbol{H}'(t) =  \boldsymbol{H}^i_{\mathrm{ti}}(t) + g\left(\boldsymbol{H}_{\mathrm{s}}(t) \right)$ where the $g(\cdot)$ is a deterministic mapping introduced by the beamforming operations. Since obtaining the ground truth for $\boldsymbol{H}_{\mathrm{s}}(t)$ is infeasible, we try to bypass this issue by using a mmWave radar at the 60~\!GHz with 2~\!GHz bandwidth~\cite{TImmWave} to collect a dataset for $\boldsymbol{H}'_{\mathrm{s}}(t)$: since the radar is designed for sensing but with system parameters similar to \name, $\boldsymbol{H}_{\mathrm{s}}(t)$ and $\boldsymbol{H}'_{\mathrm{s}}(t)$ should share high-level yet intrinsic features. 
It should be stressed that the mmWave radar is only deployed during the training stage and is surely not required by \name in runtime.
% and has excellent Tx interference isolation in its chip and circuit. 
% To further eliminate the residue Tx interference channels $ \Delta \boldsymbol{H}_{\mathrm{ti}}$ after hybrid beamforming cancellation and extract the monostatic sensing channels, 
Though we cannot directly use $\boldsymbol{H}'_{\mathrm{s}}(t)$ as ground truth for supervised learning, exploiting it to perform adversarial learning is certainly feasible.
To this end, we append an \textit{digital cancellation} neural network $\boldsymbol{C}_{\mathrm{D}}(t)$ to the above model (see Figure~\ref{fig:autodecoder}), and we train the whole model as a cGAN~(conditional Generative Adversarial Network)~\cite{mirza2014conditional}. Our model is trained to learn how to cancel Tx interference regardless of all environment inferences, so only one training is needed for each type of mmWave device.
% it only trains once before it leaves factory.} 
We denote by $\mathcal{G}$ the whole neural network constructed so far; it generates samples $\boldsymbol{x}_{g} = \mathcal{G}(\boldsymbol{z}|\boldsymbol{s})$ with $\boldsymbol{z}$ being the background Gaussian noise introduced by the signal processing/propagation pipeline (LAE). We further employ a discriminator network $\mathcal{D}(\boldsymbol{x})$ aiming to recognize if $\boldsymbol{x}$ comes from $\{\boldsymbol{H}'_{\mathrm{s}}(t)\}$.
% where  represented the probability of the real data $\boldsymbol{x}_d \sim p_{D}(\boldsymbol{x}_d)$. 
According to~\cite{mirza2014conditional}, $\mathcal{D}$ and $\mathcal{G}$ play a min-max game modeled by:
% We assume the residue Tx interference channels follow the distribution $p_{I}(\Delta \boldsymbol{H}_{\mathrm{ti}})$. Concretely, we leverage the MAP~(maximum a posteriori) to formulate the monostatic sensing channels recovery problem with the conditional distribution $p_{G}(\boldsymbol{x} | \boldsymbol{y})$ where $\boldsymbol{y} = \boldsymbol{H}(t)$:  
%
\begin{equation} \label{eq:map}
\min_{\mathcal{G}} \max_{\mathcal{D}} \mathbb{E}_{\boldsymbol{x} \sim p_{\boldsymbol{H}'_{\mathrm{s}}}}[\log \mathcal{D}(\boldsymbol{x}|\boldsymbol{s})] + \mathbb{E}_{\boldsymbol{z} \sim \mathcal{N}}[ \log(1 - \mathcal{D}(\mathcal{G}(\boldsymbol{z}|\boldsymbol{s})))], \nonumber
\end{equation}
where $\mathcal{N}$ denotes the Gaussian noise distribution.
% $L_{\mathrm{G}}(\boldsymbol{x}; \boldsymbol{y}) = -\log[ p_G(\boldsymbol{x}) ] - \log[ p_I(\boldsymbol{y} - \boldsymbol{H}_{\mathrm{s}}(t)) ]$. The above Eqn.~\eqref{eq:map} also can be allowed as the loss function $L_{\mathrm{G}}(\boldsymbol{z}; \boldsymbol{y})$ using in generator $G$ with respect to $\boldsymbol{z}$. For a given generator $G$ and a discriminator $D$, the loss function in discriminator is $L_{\mathrm{D}} = \log(1 - D(G(\boldsymbol{z})))$. 
% Different from the conventional GAN, the $\boldsymbol{z} = \boldsymbol{H}(t)$ is used to train the generator $G$ instead of noise variables sampled from a certain distribution, such as normal distribution. 
%
Essentially, we use the cGAN as a powerful non-linear filter to obtain the monostatic sensing channels by ``generating'' it out of the channel states contaminated by $\boldsymbol{H}^i_{\mathrm{ti}}(t)$.
\begin{figure}[t]
\vspace{-.5ex}
\setlength\abovecaptionskip{8pt}
\subfigure[Output (heatmap) of LAE at \textsf{1}.]{
\includegraphics[width=0.22\textwidth]{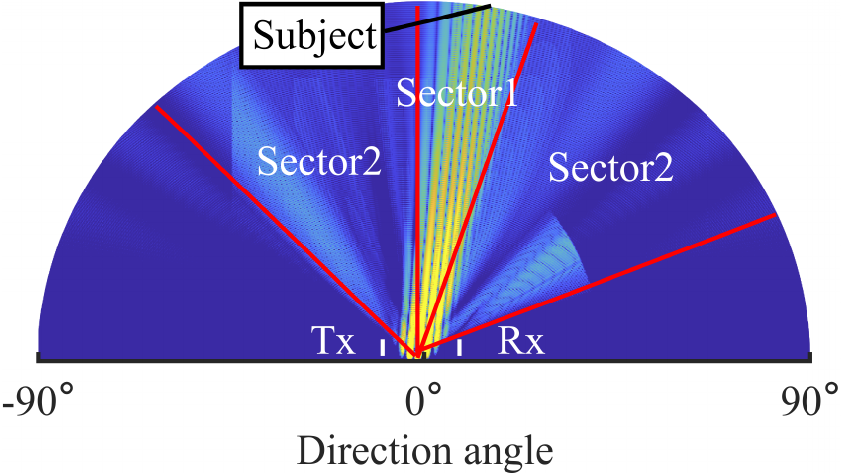}
\label{fig:beamscan}
}
%\hspace{0.5cm}
\subfigure[Output (heatmap) of $\mathcal{G}$ at \textsf{2}.]{
\centering
\includegraphics[width=0.22\textwidth]{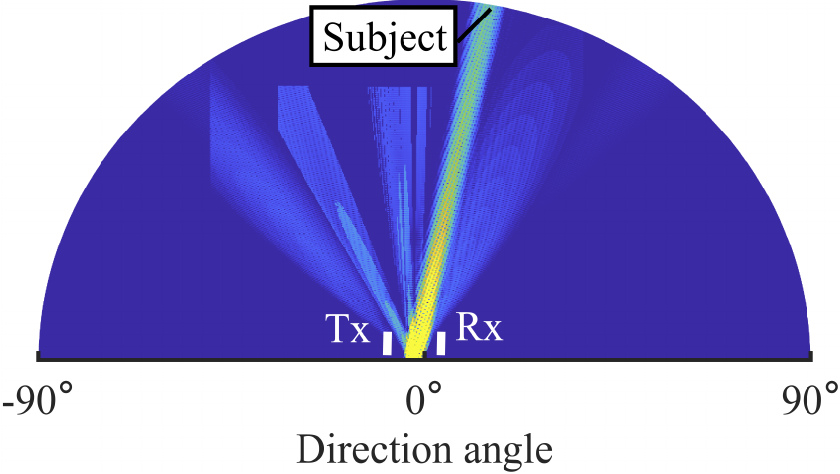}
\label{fig:TxinterCorr1}
}
\\
\subfigure[Respiration sensing results at different stages.]{
\centering
\includegraphics[width=\columnwidth]{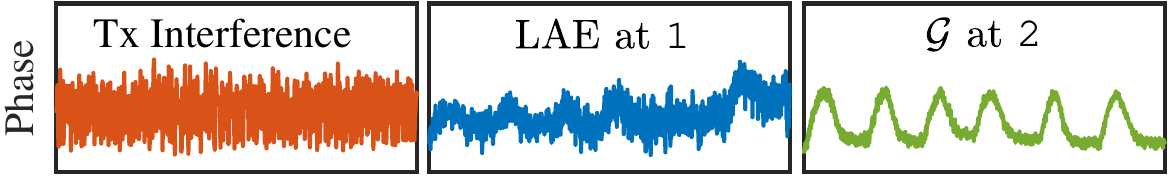}
\label{fig:TXresp}
}
\vspace{-2ex}
\caption{Beamforming heatmaps of our two-stage Tx interference cancellation (a-b) and the resulting respiration sensing outcome (c).}
\label{fig:TXintercancel}
\vspace{-2ex}
\end{figure}

% 1. Explain the difference between our model and DNN in 3.2.2 that autoencoder of our model is designed for physical mmWave beamforming. And noise is the white noise from the real world. 

% 2. mmWave radar of TI is just used in model training and not in Gemini system; our 60GHz mmWave frontend has two phased arrays and an increasing number of commercialized products with separate phased arrays.

%[zhe] decribe how to train here
% For better understanding, we summarize the learning algorithm steps in the following.
% \noindent \textit{Step~1:} We initialize the generator and discriminator with random parameters.
% \noindent \textit{Step~2:} For each training iteration, we first fix generator and then update the discriminator. We let the generator produce samples $\boldsymbol{x}_g$ with $\boldsymbol{z}$, and randomly select the mmWave radar samples $\boldsymbol{x}_d$ together. The generated and mmWave radar samples are labeled 0 and 1, respectively. In this way, the discriminator is trained to distinguish both two types of samples. 
% \noindent \textit{Step~3:} In this step, we fix the discriminator but update the generator. The generated samples are fed into discriminator and try to obtain a higher score via updating weights in generator.
% \noindent \textit{Step~4:} Finally, when the training iterations are over, we drop the discriminator and only keep the generator. The generator can extract the monostatic sensing reflection channels $\boldsymbol{H}_s(t)$. 
%

To verify the effectiveness of our design, we let a subject stand in the front of our platform with $15^{\circ}$ bearing and extract its respiration via the proposed pipeline. We first show the heatmaps of our two-stage cancellation in Figures~\ref{fig:beamscan} and~\ref{fig:TxinterCorr1}; it clearly demonstrates the gradually ``diminishing'' of $\boldsymbol{H}^i_{\mathrm{ti}}(t)$ through the pipeline. Further plotting the respiration sensing results in Figure~\ref{fig:TXresp} confirms the largely suppressed Tx interference
\footnote{While \name enables Tx interference cancellation, packet detection delays may  
introduce ranging errors. This is mitigated through an RF loopback calibration technique~\cite{guan-TMTT2021} in the RF front-end to correct the corresponding phase offsets; it processes Rx signals before ranging.}.

\vspace{-1ex}
\subsection{Beam Scheduling for mmWave ISAC} \label{ssec:sche_isac}
\vspace{-0.5ex}
% As we presented in Section~\ref{sssec:prob}, the current communication scheduling algorithms tackle either sensing or communication separately via different time slots, and the efficiency of such algorithms are too low to support mmWave ISAC. 

% The current communication scheduling does not consider the sensing physical features, such as range distance, and hence, cannot support the mmWave ISAC well. 

Since 802.11ay has a default beam scheduling scheme for MU-MIMO, 
our new scheduling scheme simply piggybacks on the existing one, aiming to integrate sensing and communication beamforming in the most efficient manner, as briefly envisioned by Figure~\ref{fig:sector}.
%
% The mmWave ISAC scheduling is more different than the communication one. As shown in the Figure~\ref{fig:mul_probe}, if we assign the subjects and UE in the same direction as more as possible to serve, the scheduling overhead will be reduced significantly. Fortunately, the MU-MIMO has been widely applied in the communication to provide multiple spatial data streaming. That gives us an opportunity to schedule MU-MIMO for both sensing and communication. Therefore, we retrofit the conventional MU-MIMO, and combine both space and time scheduling to offer better service. 
In the following, we first study how sensing subject selection differs from UE selection, 
% the selection of subjects and UEs for scheduling is studied, and 
then a scheduling algorithm is designed for efficient coverage of both sensing and communications.

\vspace{-1ex}
\subsubsection{Subject and UE Selection} \label{sssec:user_selec} 
\vspace{-0.5ex}
%1. different from comm, there is 2d resource for sensing
%2. Focus on the cover: monostatic, bistatic, communication
%3. build on the conventional comm.
%
UE selection has a default implementation in 802.11ay as MU-MIMO. In particular,
%
% such as directional sectors has existed in the current AP. In the following, the scheduling scheme for the MU-MIMO is been introduced. The scheduler in the MU-MIMO has spatial and temporal scheduling schemes. First, as described in Section~\ref{sssec:prob}, 
%
MU-MIMO starts with AP using the SSW (see Section~\ref{ssec:background}) to obtain channel states from UEs and grouping UEs with correlated channels (of similar bearing)~\cite{shen2015sieve} as a set to be covered by the same analog beam pattern. Then the bearing scheduling arranges UEs from the orthogonal (non-correlated) sets to communicate via distinct Tx chains simultaneously. For UEs covered by the same beam pattern or unable to be scheduled spatially, a temporal scheduling operating in a round-robin fashion is adopted to serve them.
%
% following UEs in the service queue. All in all, 
%
In summary, the communication-oriented UE selection operates in a \textit{single dimension} manner for bearing scheduling, leading to inefficiency in handling the conflict among pure communication, pure sensing, and simultaneous communication and sensing~(see Section~\ref{ssec:beam_schedule}) under ISAC context.

Unlike communication, sensing allows subjects with the same bearing but different ranges to be differentiated, given a sufficient range resolution. With the 2~\!GHz bandwidth of \name, the resulting centimeter-level resolution may enable many subjects to be simultaneously covered by one beam pattern, adding one more dimension for the ISAC scheduling. In addition, locations of UEs and subjects are known~(see Sections~\ref{ssec:background} and~\ref{sssec:prob}).
% Since locations of subjects are known~(see Section~\ref{sssec:prob}), the scheduler can select subjects to serve based on their direction and ranges in a two dimension manner. Consequently, to guide the scheduling design, 
Consequently, we upgrade the correlated set definition to a \textit{two-dimensional} Beam-Compatible Set~(BC-Set) in the beamforming phase; it aims to group both UEs and subjects under a single beam pattern that may serve one UE and all the covered subjects simultaneously. We plot five BC-Sets in Figure~\ref{fig:sector}; they are meant for illustrative purpose only, as realistic beam patterns (shown on the right part) can be far more irregular. 

\vspace{-1ex}
\subsubsection{Communication and Sensing Scheduling}\label{sssec:comm_multi_sta}
\vspace{-0.5ex}
To determine the BC-Sets and their corresponding beam patterns, we start from the default communication scheduling that outputs the correlated communication sets $\mathcal{U} = \{U_1, \cdots\}$~\cite{802_11ay}. As illustrated in Algorithm~\ref{alg:bcset}, our algorithm essentially executes in a greedy manner: it first adds as many subjects as possible to sets in $\mathcal{U}$ for forming the initial BC-Sets, then it constructs new BS-Sets to cover the remaining subjects.
It first gathers all positions covered by beam patterns of individual sets in $\mathcal{U}$ (line~\ref{alg:line2}). Then it performs a range-wise \textit{depth-first search} (DFS) to gather subjects that can be covered by these sets (line~\ref{alg:line3}), since initially beam patterns are often very narrow.
%
% possible along each direction bin to find all subjects in this beam pattern.  Since beam patterns used for alignment between AP and UEs are very narrow, we also search subjects in near bins of the beam pattern with covering radius $r$. 
%
If possible, it adjusts the original narrow beam pattern to trade its range for beam width, making it possible to conduct a bearing-wise breadth search for covering more subjects (line~\ref{alg:line4}). Upon upgrading all sets in $\mathcal{U}$ to BC-Sets,
% have been upgraded to BC-Sets at this stage.
% may become ``wider'' one. We continue to group subjects with UEs until all UEs are selected. 
%
the algorithm again proceeds in the greedy DFS manner in the next stage, aiming to construct new BC-Sets to cover the remaining subjects (line~\ref{alg:line7} till the end).
\RestyleAlgo{ruled}
\LinesNumbered
\begin{algorithm}[!b]
\caption{BC-Sets construction.}\label{alg:bcset}
\SetKwInOut{Input}{Input}
\SetKwInOut{Output}{Output}
\Input{
    $\mathcal{X}$: Range-bearing matrix of all entities \newline
    $\mathcal{U}$: Original communication sets for UEs  \newline
    $r$: Adjustable beam width
}
\Output{$\mathcal{B}$: BC-Sets.}

\For{$U_i \in \mathcal{U}$}{
    $\mathcal{C} \leftarrow \mathsf{beam\_pattern\_coverage}(U_i)$ \label{alg:line2} \\
    $B_i \leftarrow \mathsf{max\_depth\_range\_search}(\mathcal{C}, \mathcal{X})$ \label{alg:line3}\\
    $B_i \cup= \mathsf{breadth\_bearing\_search}(\mathcal{C}, \mathcal{X}, r)$ \label{alg:line4} \\
    % $\mathcal{D}_{q}$ = \{$D_{q}$, $\mathcal{V}$\};
}
    $\mathcal{B} = \{B_{1}, \cdots, B_{i}, \cdots\}$ \label{alg:line6}\\
    % \{ $\mathcal{D}_{1}$, \cdots, $\mathcal{D}_q$\};\\
    $\bar{\mathcal{X}} \leftarrow \mathsf{find\_non\_selected\_subjects}(B, \mathcal{X})$\label{alg:line7}\\
    %$\bar{\mathcal{X}}_{\mathrm{nearest}}$ = find_nearest_range_subjects($\bar{\mathcal{X}}$);\label{alg:line8}\\
%
\While{$\bar{\mathcal{X}} \ne \emptyset$}{
    %$\mathcal{C}$ = beam_pattern_coverage(q);\label{alg:line10}\\
    $[\mathcal{C}, B] \leftarrow \mathsf{max\_depth\_range\_search}(\bar{\mathcal{X}})$ \\
    $B \cup= \mathsf{breadth\_bearing\_search}(\mathcal{C}, \mathcal{X}, r)$ \label{alg:line4} \\
    $\bar{\mathcal{X}} \leftarrow \mathsf{find\_non\_selected\_subjects}(B, \bar{\mathcal{X})}$ \\
    $\mathcal{B}~\cup = \{B\}$
}   
% \vspace{-2ex}
\end{algorithm}
%
% are used to identify all non-selected subjects, and find some subjects with the closest range in them. We repeat the similar steps as the first phase to group these subjects.
% The communication sets $\mathcal{U}_{\mathrm{c}}$ are determined by the communication scheduler, and hence, the $\mathcal{U}_{\mathrm{c}}$ is used as the initial sets for us. 
% Since the most of analog beam patterns are very narrow for better alignment in communication, we perform a depth search at the first to select the subjects close to the beam patterns of $\mathcal{U}_{\mathrm{c}}$. 
% For instance, as shown in the Figure~\ref{fig:sector},  the subjects with different range bins are selected in the beam pattern of a UE as the BC-Set~(red color) via depth search. After that, there are residue subjects that are not selected, and they are clustered in different sets with a hyper-parameter covering radius $\mathcal{R}$ according to their direction and range bins. In this way, all BC-Sets are obtained, and report to the scheduler. 
%
\begin{figure}[t]
%\vspace{-1ex}
   \setlength\abovecaptionskip{8pt}
   \centering
   \includegraphics[width=.92\columnwidth]{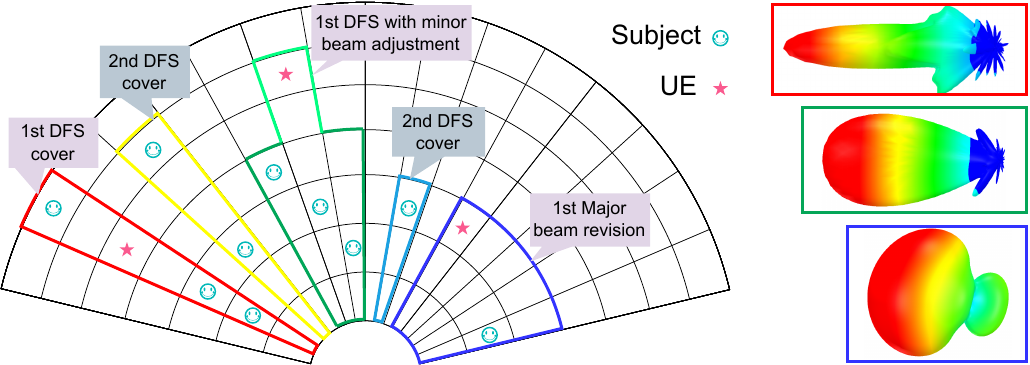}
   % \vspace{-0.3cm}
   \caption{{BC-Sets for ISAC-oriented beam scheduling.}}
   \label{fig:sector}
   \vspace{-1.5em}
\end{figure}

What Algorithm~\ref{alg:bcset} outputs are only (logical) BC-Sets; they need to be ``translated'' into (physical) beam patterns. Essentially, the beam patterns should be shaped by both the distribution of UEs/subjects covered by the BC-Sets and the capability of the phased array~\cite{gu2021packaging}. The procedure of shaping a beam pattern follows two basic principles: i) adopting wider beams to trade range for width coverage and ii) adjusting the direction of the main beam to balance the coverage among multiple entities. Since the resulting patterns have not taken into account the beam nulling requirements (see Section~\ref{ssec:mono_sens}), the corresponding beamformers are used as the initial setting for the problem~\eqref{eq:m-ssnr} (hence the LAE) to derive proper cancellation schemes for individual BC-Sets. 
% \rev{According to BC-Sets scheduling, the AP performs beamforming~(see Section~\ref{ssec:background}).}
%
Finally, we analyze the complexity of Algorithm~\ref{alg:bcset} to demonstrate its efficiency for real-time execution. 
%
% We denote the total number of UEs and subjects as $K_{\mathrm{U}}$ and $K_{\mathrm{S}}$~(where $K = K_{\mathrm{U}} + K_{\mathrm{S}}$).
Let the number of UEs be $K$ and that for range (resp. bearing) bins be $K_{\mathrm{R}}$ (resp. $K_{\mathrm{D}}$), the complexity is $O(K K_{\mathrm{D}}  K_{\mathrm{R}}  )$, 
% which is often 
often up to the scale of several hundreds only. 
\begin{figure}[b]
\vspace{-2ex}
\setlength\abovecaptionskip{0pt}
\subfigure[One Tx chain.]{
\includegraphics[width=0.22\textwidth]{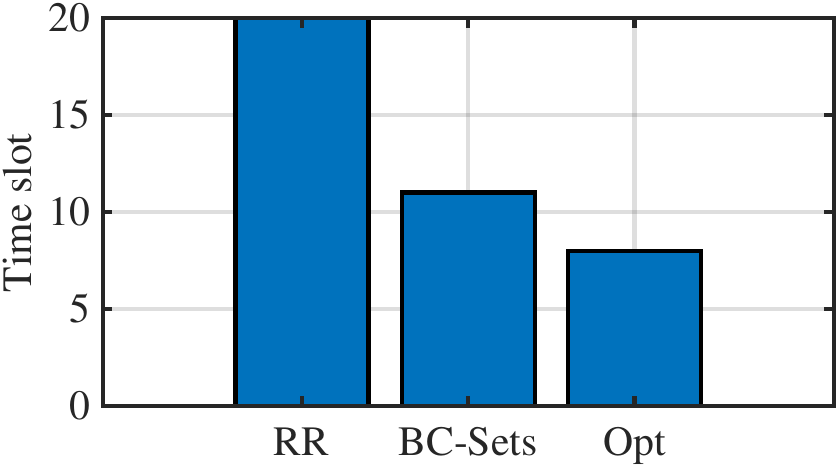}
\label{fig:bcset11}
}
\hfill
\subfigure[Two Tx chains.]{
\centering
\includegraphics[width=0.22\textwidth]{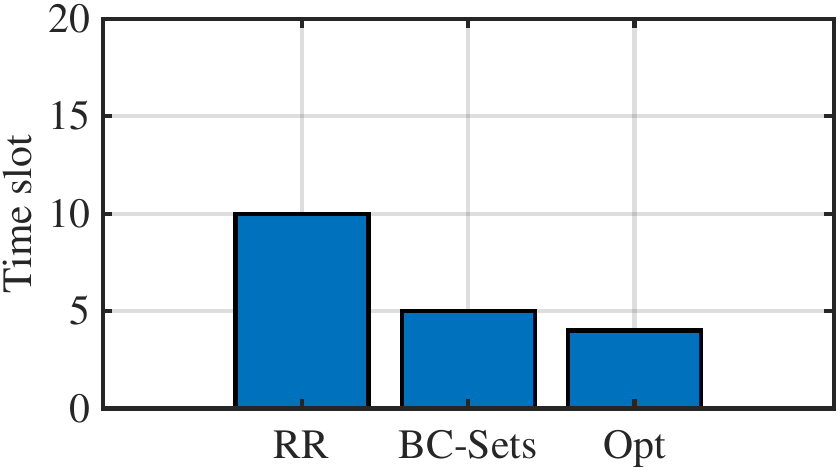}
\label{fig:bcset22}
}
%\vspace{-1.5ex}
\caption{Average time spans of round-robin (RR), BC-Sets (scheduling) and the optimal solution (Opt).}
\label{fig:sche_overhead}
\vspace{-.5ex}
\end{figure}

We may walk through the algorithm using the BC-Sets in Figure~\ref{fig:sector} as examples. During the 1st stage, the three UE sets (red, green, and blue) are sequentially upgraded to form BC-Sets by adding in more subjects. Then the remaining three subjects are covered by two new BC-Sets (yellow and cyan). While the red and yellow BC-Sets can be covered by the default (narrow) beam sectors, others should be reshaped: in particular, both principles apply to the blue one.
%When the BC-Sets have been constructed, according to different BC-Sets, the beam patterns need to adapt their shapes following above two basic principles. For example, due to wider main beam width of green BC-Set, original beam pattern only for UE enlarges its coverage
% the AP generates new multiple Tx beam patterns jointly optimized $\{\boldsymbol{H}_{\mathrm{s}}(t), \boldsymbol{H}_{\mathrm{c}}(t), \boldsymbol{H}_{\mathrm{ms}}(t)\}$. However, their beamformers $\boldsymbol{W}_{\mathrm{AB}}^{\mathrm{Tx}} \boldsymbol{W}_{\mathrm{DB}}^{\mathrm{Tx}}$ do not consider 
%
%
% Consequently, we need to fine-tune the beamformers for balancing all channels in $\boldsymbol{H}(t)$. Since the $\boldsymbol{W}_{\mathrm{DB}}^{\mathrm{Tx}}$ is determined by the default MU-MIMO illustrated in Section~\ref{sssec:user_selec}, we focus on fine-tuning  $\boldsymbol{W}_{\mathrm{AB}}^{\mathrm{Tx}}$. We denote $\boldsymbol{W}_{\mathrm{AB}}^{\mathrm{Tx,null}}$ and $\boldsymbol{W}_{\mathrm{AB}}^{\mathrm{Tx,bc}}$ as the analog beamformers determined by beam nulling and BC-Sets, respectively. We minimize the problem 
% $\min_{\boldsymbol{W}_{\mathrm{AB}}^{\mathrm{Tx,mid}}} $ 
% MSE($||\boldsymbol{W}_{\mathrm{AB}}^{\mathrm{Tx,null}} -  \boldsymbol{W}_{\mathrm{AB}}^{\mathrm{Tx,mid}} ||_{F} $, $ ||\boldsymbol{W}_{\mathrm{AB}}^{\mathrm{Tx,mid}} - \boldsymbol{W}_{\mathrm{AB}}^{\mathrm{Tx,bc}} ||_F$), and solve it to obtain $\boldsymbol{W}_{\mathrm{AB}}^{\mathrm{Tx,mid}}$ as fine-tuned analog beamformers.
%
We also perform a trace-driven emulation to evaluate the advantages of BC-Sets. We set 6 UEs and 14 subjects in our experiment, and randomize their positions in each experiment. We run three algorithms each for 100 trials, and show the average time spans for serving all UEs and subjects in Figure~\ref{fig:sche_overhead}. Apparently, our scheduling requires only half of RR's time span and goes close to the optimal solution. 

\vspace{-1.5ex}
\subsection{Exploiting Sensing Diversity} \label{ssec:hybrid_sensing}
\vspace{-.5ex}
The scheduling algorithm presented in Section~\ref{ssec:sche_isac} has actually implicated two steps. On one hand, this scheduling takes place not only on the AP side but also on the UE side, albeit with less powerful beam fine-tuning ability. On the other hand, whereas covering monostatic sensing subjects can be determined in a unilateral manner, conducting bistatic sensing demands collaborative beamforming on both sides, similar to the default (bistatic) communication scheduling. 
As the locations of all subjects are estimated in advance (see Section~\ref{sssec:prob}), all these can be achieved naturally by Algorithm~\ref{alg:bcset}. Therefore, the question now is how to exploit the sensing diversity introduced by different sensing modalities, 
% Leaving the sensing information passing/exchanging to the protocol-level discussions in Section~\ref{ssec:protocol_comp}, we hereby 
and our answer is the following unified estimation framework at signal processing level.
% , and the information passing/exchanging protocol-level modifications.}

\begin{figure}[h]
  \setlength\abovecaptionskip{8pt}
  \centering
  \includegraphics[width=1\columnwidth]{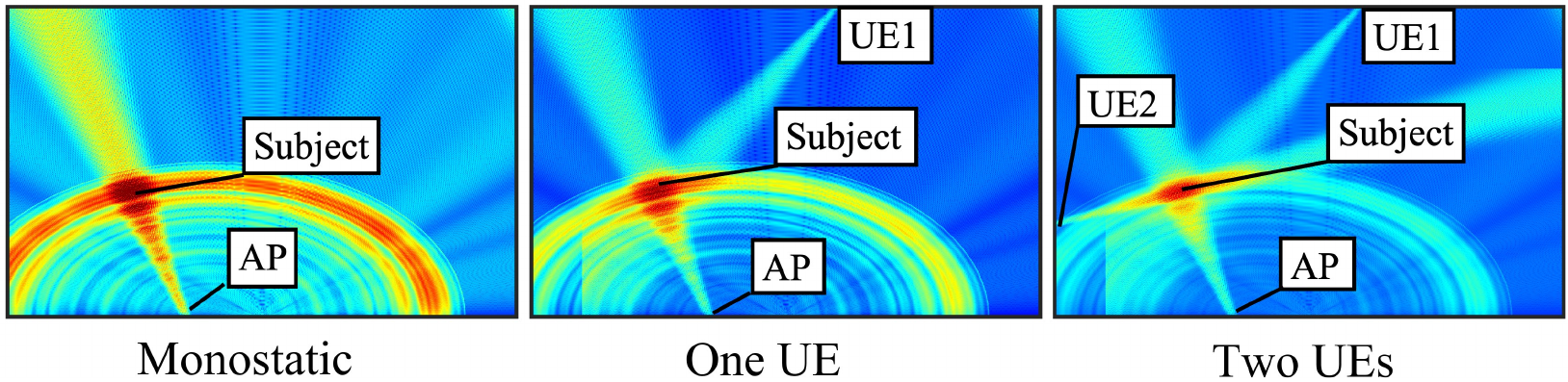}
 \caption{Improving estimation precision via joint monostatic and multi-static sensing.}
 % (Left)~single monostatic sensing modality, (Center)~monostatic and bistatic modalities and (Right)~monostatic and multi-static sensing modalities.
    \label{fig:sensing_diversity}
    \vspace{-1.5ex}
\end{figure}

One of the major reasons that motivates us to consider full-fledged ISAC is the deficit of multi-static sensing in lack of synchronizations among different parties~\cite{isacot}. As a result, multi-static sensing can only be used to sense motion-induced information that incurs channel variations, whereas sensing (quasi-)static subjects can be fully achieved by monostatic sensing: for example, locating a subject can be done by AP estimating range/bearing and refined by the same estimations from UEs. 
Nevertheless, in motion tracking scenarios for both micro-motion~\cite{movifi} and macro-motion~\cite{jiang2018towards}, joint estimation with diversified sensing results can only lead to improved accuracy~\cite{fleury1999channel}. Instead of dwelling on the design of estimation algorithms, we simply leverage existing estimation frameworks for this purpose. On the one hand, micro-motion estimation is taken care by SAGE~\cite{fleury1999channel}, which requires a subject to remain quasi-static. On the other hand, macro-motion estimation (a.k.a motion tracking) is handled by the well-known extended Kalman filter~\cite{kalman}.  We demonstrate how sensing diversity helps to improve precision of subject location estimation (via SAGE) in Figure~\ref{fig:sensing_diversity}: from left to right, the ``hot spot'' shrinks as the number of sensing modalities (hence the amount of beam intersections) increases, indicating a higher estimation precision. 
%
% The left figure represents single monostatic sensing modality with single AP that circle~(range) and ray~(bearing) are intersected together to point out subject. The central figure depicts the AP with an additional UE1 being bistatic mode, and even without range information, the ray of UE1 can help to refine motion of subject via combining the monostatic snesing. Similar to the central figure, the third figure shows the more sensing modalities we have, the better performance we achieve.
%

\vspace{-0.8ex}
\section{Implementation} 
\begin{figure}[t]
  \setlength\abovecaptionskip{8pt}
  \centering
  \includegraphics[width=.78\columnwidth]{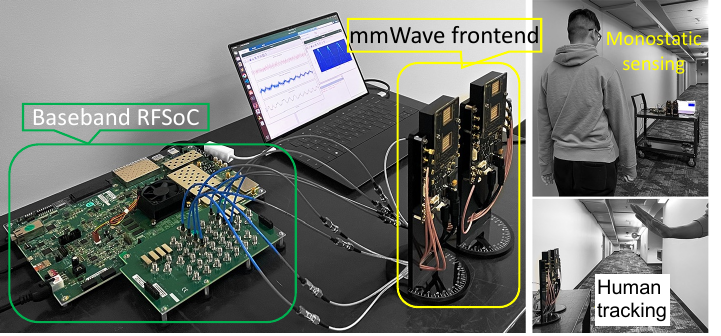}
 \caption{Part of the implementation and evaluation setup with mmWave frontend and baseband RFSoC.}
    \label{fig:evalsteup}
    \vspace{-1.5em}
\end{figure}

%\vspace{-0.5ex}
\label{sec:impl}
% In this section, 
We hereby elaborate on the implementation of 
% describe the steps taken in implementing and building the 
\name, particularly explaining how to configure \name for carrying out the design objectives outlined in Section~\ref{sec:design}. The implementation follows IEEE 802.11ay standard operating at the unlicensed 60~\!GHz frequency band, as demonstrated in Figure~\ref{fig:evalsteup}. We omit the illustration for the UE part, as it involves similar components but with only one mmWave frontend. 
%
% band shown far left in Figure~\ref{fig:evalsteup}.
% and provides the capability to estimate the CIR for various beam patterns.
%

\vspace{-1ex}
\paragraph{Hardware Platform} 
\name has three hardware components: \mm front-end module, baseband processing module, and high-performance processor module.
% shown in Figure~\ref{fig:evalsteup} (left-and side). 
For the RF front-end, \name adopts the EVK06003 development kits from Sivers Wireless~\cite{sivers}; each kit has two 16-element phased arrays\footnote{Since many commercial transceiver chips of mmWave offer separated Tx and Rx phased arrays~\cite{sivers,qualcommwigig}, \name is equipped with two such arrays.}
% (see Appendix~\ref{append:b} for more details).}
The baseband processing and high-performance processor modules are both realized upon the Xilinx RFSoC ZCU208 development board~\cite{xilinxzcu208}; it offers a multitude of advanced features and capabilities, including i) AD/DA converters for baseband sampling with a rate up to 4~\!GHz, ii) DDR memory providing ample storage space for processed data, and iii) multi-core ARM processors and high-end Ultrascale FPGA offering substantial computational power.

\vspace{-1ex}
\paragraph{Software Components}
%
% The software component of \name is integral in the overall software functionality of the system. It 
The software of \name is responsible for a variety of tasks, including beamforming, interference cancellation~(Section~\ref{ssec:mono_sens}), beamform scheduling~(Section~\ref{ssec:sche_isac}), data streaming to and from a PC controller. We implement above tasks using Verilog and C/C++, and compile them as a firmware for Xilinx ZYNQ RFSoC. We also implement a Matlab interface 
% for controlling the \name 
to pull the data from the RFSoC into the PC. The sensing algorithms, except for the deep neural module, are implemented in Matlab.

% This part is comprised of the Xilinx ZYNQ RFSoC firmware, programs for controlling the 60GHz mmWave front-end, and a Matlab interface for controlling the \name platform and for exporting data. The combination of these elements allows for a comprehensive and efficient software solution to the various tasks at hand.

% \emph{\textbf{IEEE 802.11ad/ay compliant:}} \name has been engineered to operate as a full-duplex transceiver, enabling it to transmit and receive at the same time. The transmitter block can transmit full-bandwidth frames in compliance with the IEEE 802.11ad~\cite{802_11ad} standard, and it can do so independently of the receiver section. This is accomplished by directly streaming IQ samples from the dedicated DDR memory to the Tx data path, which results in full-duplex performance that is compliant with the IEEE 802.11ad standard.

\vspace{-1ex}
\paragraph{Deep Neural Network} 
Our deep neural network is built upon PyTorch~\cite{pytorch} platform using Python~3.7, and an mmWave radar~\cite{TImmWave} is adopted
% and DCA1000EVM data capture adapter~\cite{DCA1000EVM} 
to act as both a baseline and ground truth sensor. We synchronize the clocks of the radar and \name (both driven by a PC) via precision time protocol~\cite{ptp}, hence aligning their starting time to $\mu$s level. 
% An decoder layer of LAE adopts CNN kernel of size $3\times3$, and its layer has a stride of 1, padding of 0, and dilation of 1. All outputs with with the kernel do not use any activated function.  The encoder layer also uses the same kernels as the decoder, but they are connected sequentially. 
The encoder/decoder of the LAE are actually formed by beamformers, so we only need to construct $\boldsymbol{C}_{\mathrm{D}}(t)$ as 
a 5-layer perceptron and the discriminator with three CNN layers whose input size ($2000 \times 1$) matches the output data of the radar. The batch size is set to 64 for training, and the model is optimized using the Adam optimizer with a learning rate of 0.001. 
We further quantize the neural network weights~\cite{HAQ_CVPR2019} to control the 4-bit phase and amplitude via AWV look-up table of mmWave frontend.
%
% The weights is  integerized and fit to codebook through Antenna Weight Vector Look-up Table (AWV LUT) built in front end.

\vspace{-0.5ex}
\section{Experiment Evaluations}
\label{sec:eval}
\vspace{-.5ex}
We now perform three sets of experiments to verify the major functions of \name, namely Tx interference cancellation for monostatic sensing, beam scheduling for joint sensing and communications, and sensing diversity exploitation. Part of the experiment setup is depicted in Figure~\ref{fig:evalsteup}, but other necessary details shall be provided later. Our experiments have strictly followed the IRB of our institutes.

\begin{figure}[b]
\vspace{-1.5em}
  \setlength\abovecaptionskip{0pt}
  \centering
  \includegraphics[width=0.95\columnwidth]{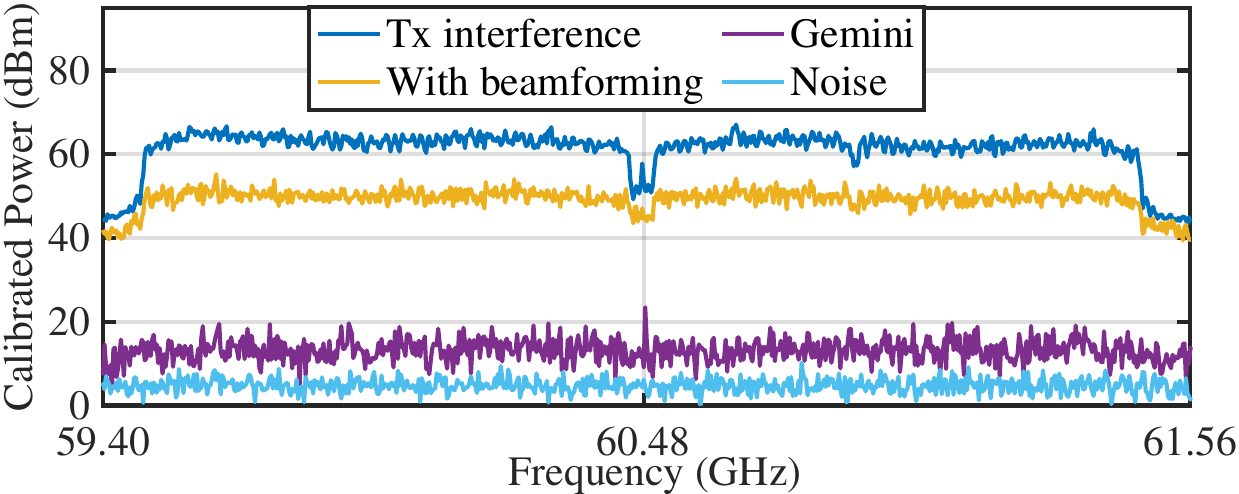}
 \caption{Power spectrum of the received baseband signal after two-stage Tx interference cancellation.}
    \label{fig:TXCancelBench}
    \vspace{-.5ex}
\end{figure}

% \needrev{1. Error bar; 2. legend is too long: TI, \name, Wo/c. 3. box on}.
\begin{figure*}[t]
\centering
\vspace{-1.5ex}
\setlength\abovecaptionskip{8pt}
\subfigure[Ranging.]{
\includegraphics[width=0.32\textwidth]{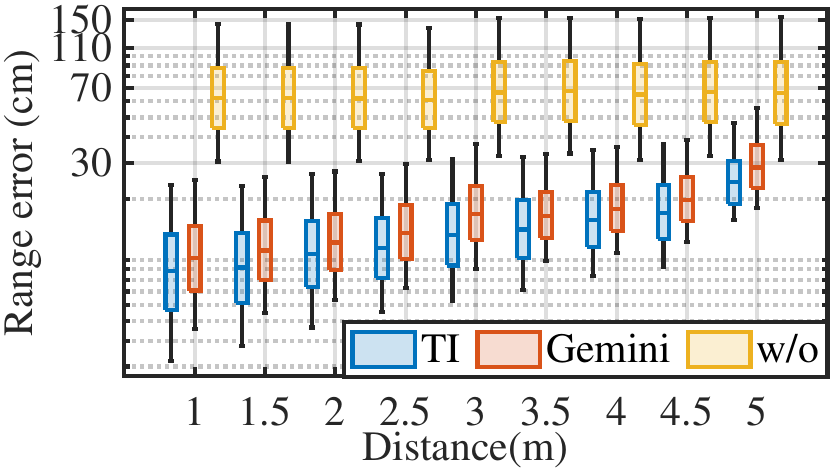}
\label{fig:RangeBench}
}
%\needrev{ 1.Error bar 2.  Wo/c has more errors, about 30\% relative error 3. different angle, such as -60 -30, 0, 30, 60}.
\subfigure[Angle of arrival (AoA).]{
\centering
\includegraphics[width=0.32\textwidth]{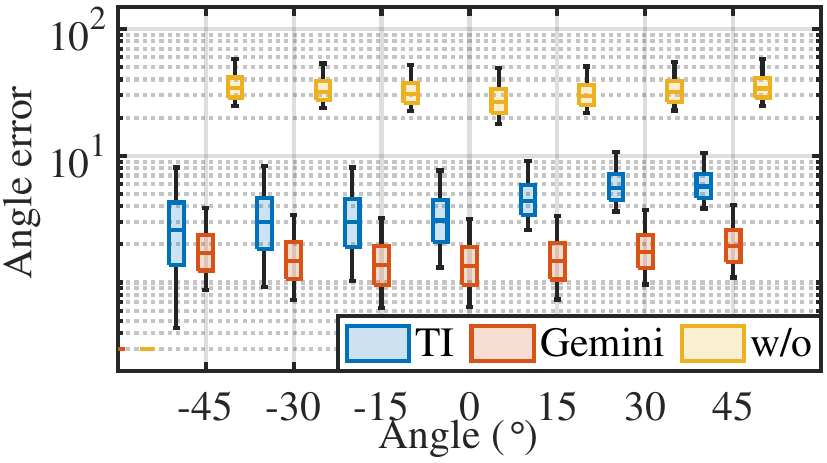}
\label{fig:aoaBench}
}
%
%\needrev{1. error bar. 2. only two error bars: one is  Wo/c; the other is \name. The TI radar is used as ground-truth device. 3.  Wo/c has at least 30\% relative error than \name} 
\subfigure[Motion sensing.]{
\centering
\includegraphics[width=0.32\textwidth]{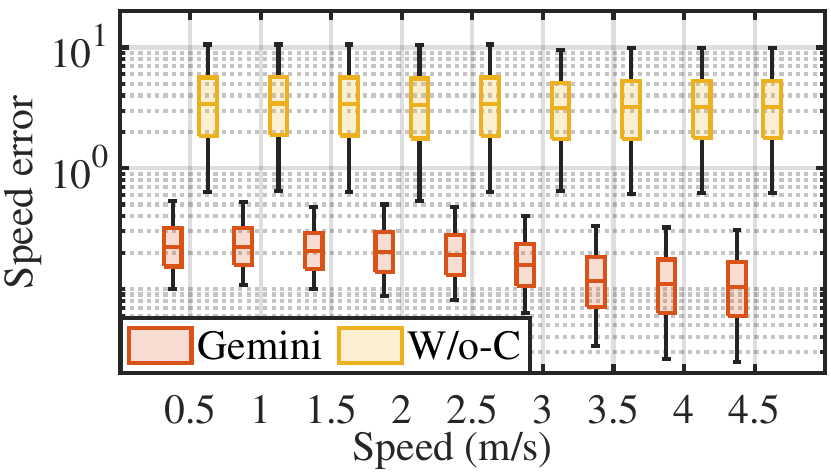}
\label{fig:motionBench}
}
\vspace{-2.5ex}
\caption{Monostatic sensing performance. ``TI'' and ``W/o-C'' denote the TI mmWave cascaded radar and \name without Tx interference cancellation, respectively. TI disappears from (c) as it is used as the ground truth collector.}
\label{fig:mono_sens_perfor}
\vspace{-2ex}
\end{figure*}

\vspace{-1ex}
\subsection{Monostatic Sensing}
\vspace{-.5ex}
We report the outcome of Tx interference cancellation and the consequent monostatic sensing performance of \name. The experiments involve range, bearing, and speed estimations of a static or moving object (a metal block held by a person).
% are conducted to evaluate the fundamental functionality and involved 10 subjects. 
The ground truths for range and bearing are measured by Intel RealSense LiDAR~\cite{realsense-L515}, while that for speed is provided by the TI radar~\cite{TImmWave}. We adopt absolute error as the evaluation metric.
%
% The results showcase the performance of interference cancellation, range estimation, Angle-of-Arrival (AoA) estimation, and velocity estimation.
\vspace{-1ex}
\subsubsection{Tx Interference Cancellation}
\vspace{-.5ex}
%
% The purpose of our study is to examine how Tx interference cancellation~(see Section~\ref{ssec:mono_sens}) performs and its impact. 
We begin by quantifying the interference cancellation ability at different stages. We set the Tx power to 40\!~dBm
% Tx power to evaluate the performance of Tx interference cancellation and 
and observe the signal strength at different stages indicated in Figure~\ref{fig:autodecoder}. The results shown in Figure~\ref{fig:TXCancelBench} confirm that LAE reduces Tx interference by about 39\!~dB, and deep denoising further suppresses it by 27\!~dB. Overall, the total cancellation is approximately 66\!~dB, and the remaining Tx-interference power goes very close to the noise floor. Apparently, the monostatic reflection channels $\boldsymbol{H}_{\mathrm{s}}(t)$ has been distilled from $\boldsymbol{H}(t)$ in Eqn.~\eqref{eq:csi}.

\vspace{-1ex}
\subsubsection{Range}
\vspace{-.5ex}
%Monostatic v.s bistatic : aoa, velocity
% 
The ability to ranging is a fundamental and crucial sensing feature, but it can only be accomplished in the monostatic mode~\cite{isacot}. We study the ranging performance with/without Tx interference cancellation, 
%
% for each trial, we ask distinct subjects to stand 
given the subject staying statically in the range of $[1, 5]$~\!m with a step size of 0.5~\!m from \name,
% in a corridor, 
and \name performing ranging via CIR. We use the TI radar as a baseline by co-locating it
% put it in same position 
with \name. The range errors shown in Figure~\ref{fig:RangeBench} confirm that \name can obtain comparable performance to the radar,
% using Tx interference cancellation. 
but much higher median range errors up to 30~\!cm are introduced without the Tx interference cancellation. The range errors of both TI radar and \name slightly grow as the distance increases, potentially because the adopted beam pattern covers more background clutters at further distances and multipath reflections from them affect the ranging accuracy. Moreover,  reflected (sensing) signals attenuate in distance and hence also result in degraded performance.

% \begin{figure}[t]
%   \setlength\abovecaptionskip{8pt}
%   \centering
%   \includegraphics[width=0.8\columnwidth]{figures/RangeBenchmark.pdf}
%  \caption{\needrev{The power spectrum of the received baseband Tx signals for without cancellation, beamforming, as well as for the deep denoising algorithm.} \needrev{1. Error bar; 2. legend is too long: TI,  ,W/c. 3. boxing}}
%     \label{fig:RangeBench}
%     \vspace{-1ex}
% \end{figure}
\vspace{-1ex}
\subsubsection{Bearing}
\vspace{-.5ex}
%Monostatic
To study another fundamental sensing function, we leverage the phased arrays equipped with both \name and the TI cascaded radar~\cite{TImmWave}
% mmWave radio 
to estimate object bearing (or AoA). We let the object stay within a bearing of $[-45^{\circ}, 45^{\circ}]$ and vary with a step size of $15^{\circ}$. 
% and the phased array is used to scan each of them via the minimum narrow beam pattern with \needrev{$1.5^{\circ}$}. Apparently, 
As shown in Figure~\ref{fig:aoaBench}, \name achieves better performance than that of radar because \name has a much more powerful phased array than that of radar. Again, Tx interference significantly degrades the bearing estimation performance if not properly handled. 
% leading to large estimation error. 

% \begin{figure}[t]
%   \setlength\abovecaptionskip{8pt}
%   \centering
%   \includegraphics[width=0.8\columnwidth]{figures/RangeBenchmark.pdf}
%  \caption{AoA estimation error. \needrev{   }}
%     \label{fig:RangeBench}
%     \vspace{-1ex}
% \end{figure}

\vspace{-1ex}
\subsubsection{Speed}
%\vspace{-1ex}
%
The motion sensing is also crucial 
% very important for \name, and in order to verify the stated 
to the capability of \name, so we conducted tests where the object moves at a varying speed ranging from 0.5\!~m/s to 3.0\!~m/s within a 3 $\times$ 20\!~$\text{m}^2$ corridor. Since the LiDAR cannot monitor motion, we have to change the role of the TI radar from baseline to measuring the ground truth. The evaluation results shown in Figure~\ref{fig:motionBench} clearly demonstrate that \name achieves a much lower speed estimation error with the enhancement offered by the Tx interference cancellation. 
% \needrev{This experiment also justifies the observation we depicted in Section~\ref{ssec:nfs}.}

\vspace{-1ex}
\paragraph{Remark:} During all above experiments, we have a communication session going on between the AP and UE. Since monostatic sensing takes place only on the AP side and it simply piggybacks on the Tx signals, the communication throughput 
is not affected at all.
%
% demonstrate the capability of  basic functionality~(range, direction and motion) for monostatic sensing. Moreover, throughput in those experiments is not impacted, due to AP and UE beam alignment. Tx interference degrade all of sensing functionality, but \name employs cancellation~(designed in Section~\ref{ssec:mono_sens}) to suppress it, and obtain  comparable performance with TI mmWave radar. 

% \begin{figure}[t]
%   \setlength\abovecaptionskip{8pt}
%   \centering
%   \includegraphics[width=0.8\columnwidth]{figures/RangeBenchmark.pdf}
%  \caption{Velocity estimation error.}
%     \label{fig:RangeBench}
%     \vspace{-1ex}
% \end{figure}

% \subsubsection{Monostatic sensing v.s. Bistatic sensing}

\begin{figure}[b]
    \vspace{-1em}
  \setlength\abovecaptionskip{8pt}
  \centering
  \includegraphics[width=0.88\columnwidth]{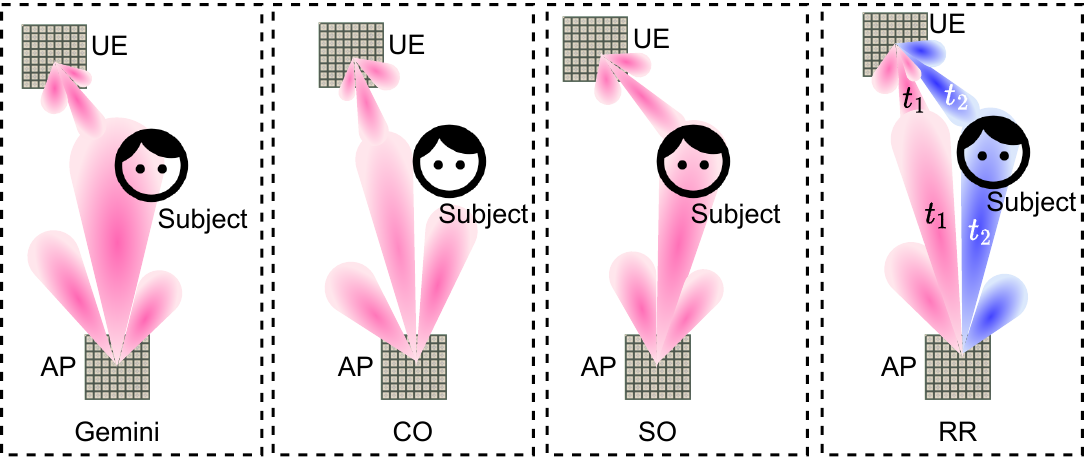}
 \caption{An illustration of the experiment setup.}
    \label{fig:exp_dep}
    \vspace{-.4em}
\end{figure}

\vspace{-0.5ex}
\subsection{Beam Scheduling for ISAC}
\vspace{-0.5ex}
%
% Our scheduling algorithm overlays current scheduling of MAC layer, and its efficiency has been described in Section~\ref{ssec:sche_isac}. 
One may expect a full-scale evaluation of our beam scheduling algorithm with many subjects and UEs involved simultaneously. Our experimental setup, depicted in Figure~\ref{fig:exp_dep}, is straightforward, involving one subject and one UE. Specifically, we demonstrate the simultaneous tracking of two subjects with two UEs in human tracking experiments to illustrate the potential for involving additional UEs.
%%but we actually adopt only a simple setup as illustrated in Figure~\ref{fig:exp_dep}, with one subject and one UE instead.
%
On one hand, 
%involving multiple subjects is not necessary, as 
our scheduling algorithm excels in dispatching BC-Sets to cover subjects with compatible demands and hence all BC-Sets (or all subjects covered by a BC-Set) are similarly served by the AP.
% as far as the AP has available spatial and temporal resources. 
Consequently, evaluating the performance with one subject is sufficiently representative.
% and evaluating communication and sensing performance in all BC-Sets are not necessary. Moreover, 
On the other hand, 
serving multiple UEs bears no difference from serving one UE, as multiple UEs would inevitably need to be served by distinct Tx chains or in different time slots (see the media access described in Section~\ref{ssec:background}). 
Also, 
due to the extremely high cost of our high-performance hardware (mmWave frontend~\cite{sivers} each costs over \$~\!3300 and RFSoC~\cite{xilinxzcu208} costs \$~\!15000), we cannot afford to support a lot of UEs.
% such a large number of mmWave platforms for evaluating all BC-Sets. 
% but a trace-driven emulation results has been shown in Figure~\ref{fig:sche_overhead}. 
%
Therefore, we 
% study communication and sensing performance in a single BC-Set, and also 
believe our evaluation scenario does produce results with practical significance.
%\footnote{\newrev{Since experiment evaluation on large-scale deployment cannot be achieved, we leverage a simulation to demonstrate performance of our beam scheduling algorithm for large-scale networks in Appendix~\ref{append:c}.}}
% can be generalized to any BC-Set. 
%
In the following, we evaluate our scheduling algorithm in three sensing applications, namely respiration monitoring, point cloud of human pose, and human tracking, along with a communication session. We consider three scheduling baselines: sensing only~(SO), communication only~(CO)~\cite{SideLobe-UbiComp23}, and round-robin~(RR), 
% to compare with our algorithm. In addition, we also utilize 
while also adopting the TI radar as a sensing baseline. 

% \name vs
% baselines: 1.only for communication (OC) 2. only for sensing(OS) 3. round-robin(RR)

\vspace{-1.3ex}
\subsubsection{Respiration Monitoring} \label{sssec:respiration}
\begin{figure}[t]
\vspace{-0.5ex}
\setlength\abovecaptionskip{0pt}
\subfigure[{Throughput.}]{
\includegraphics[width=0.22\textwidth]{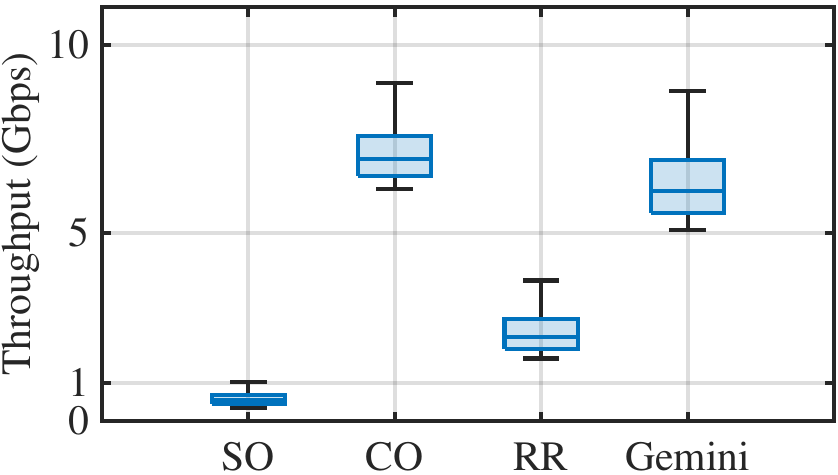}
\label{fig:bcset11}
}
%\hspace{0.5cm}
\subfigure[Breath rate error.]{
\centering
\includegraphics[width=0.22\textwidth]{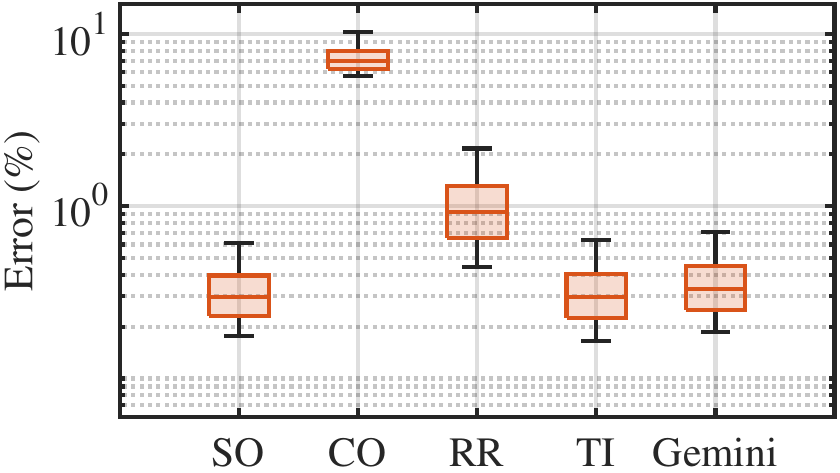}
\label{fig:bcset22}
}
%\vspace{-1ex}
\caption{Scheduling for simultaneous communications and respiration monitoring.} 
%\needrev{ there are 5 error bars including OC, OS, RR , TI and \name for both figures. How to do: 1. we use the 100\% sampling rate to sample breath motion for OS, and  100\% sampling rate throughput for OC. 2. 80\% performance of OS and OC in both throughput and breath for \name. 3. about 30\% sensing performance of OS for OC. 4. about 30\% throughput of OC for OS. 4. 50\% throughput of OC and 50\% sensing performance of OS for RR.}
\label{fig:sche_breath}
\vspace{-1.5ex}
\end{figure}

%\vspace{-0.5ex}
%
This experiment takes the subject's respiration as the sensing target, aiming to accurately estimate the breath rate. The results plotted in Figure~\ref{fig:sche_breath} showcase the obtained throughput against corresponding sensing accuracy. We may observe that, while \name achieves a sensing accuracy almost the same as that for both SO and TI radar (which is much better than that for CO and RR), its throughput is only marginally lower than that of CO (but still much higher than that for SO or RR). In fact, for continuous yet spatially coarse-grained sensing applications, \name can always obtain nearly perfect sensing performance and barely sacrifice throughput.

% \begin{figure}[t]
% \vspace{-2ex}
% \setlength\abovecaptionskip{8pt}
% \subfigure[Throughput.]{
% \includegraphics[width=0.22\textwidth]{figures/bcset11.pdf}
% \label{fig:bcset11}
% }
% %\hspace{0.5cm}
% \subfigure[Breath rate error.]{
% \centering
% \includegraphics[width=0.22\textwidth]{figures/bcset22.pdf}
% \label{fig:bcset22}
% }
% \vspace{-2ex}
% \caption{multi-static sensing}
% \label{fig:sche_overhead}
% \vspace{-1ex}
% \end{figure}

\begin{figure}[!b]
\vspace{-2ex}
\setlength\abovecaptionskip{0pt}
\subfigure[Throughput.]{
\includegraphics[width=0.22\textwidth]{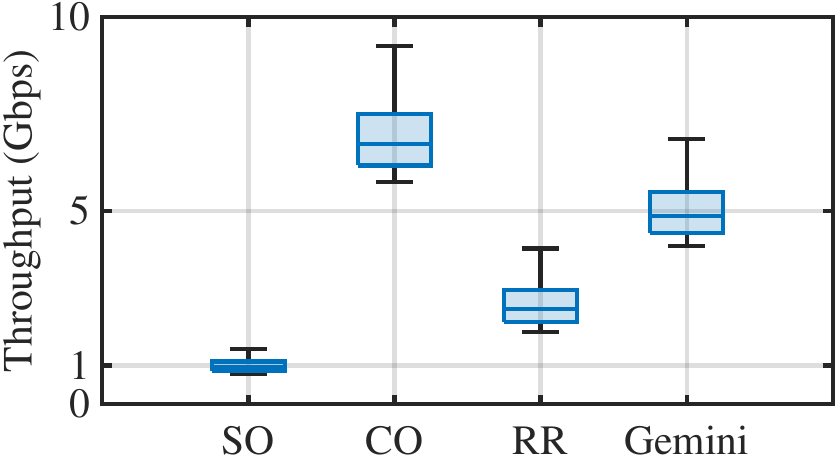}
\label{fig:pointcloudthroughput}
}
%\hspace{0.5cm}
\subfigure[\name.]{
\centering
\includegraphics[width=0.22\textwidth]{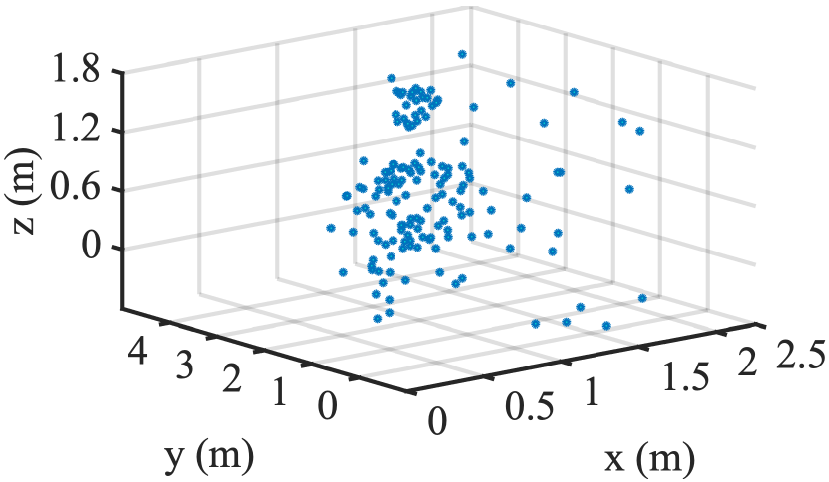}
\label{fig:pointcloudGemini}
}
\subfigure[Sensing only (SO).]{
\centering
\includegraphics[width=0.22\textwidth]{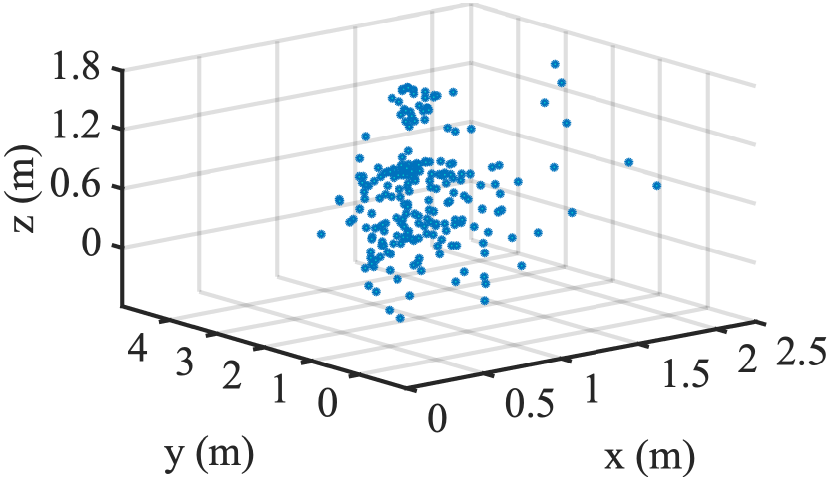}
\label{fig:pointcloudSO}
}
\subfigure[Communication only (CO).]{
\centering
\includegraphics[width=0.22\textwidth]{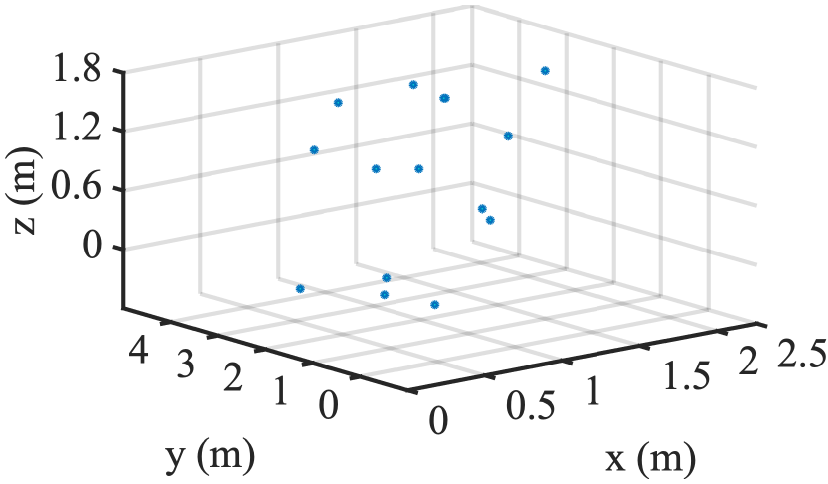}
\label{fig:pointcloudCO}
}
\subfigure[Round-robin (RR).]{
\centering
\includegraphics[width=0.22\textwidth]{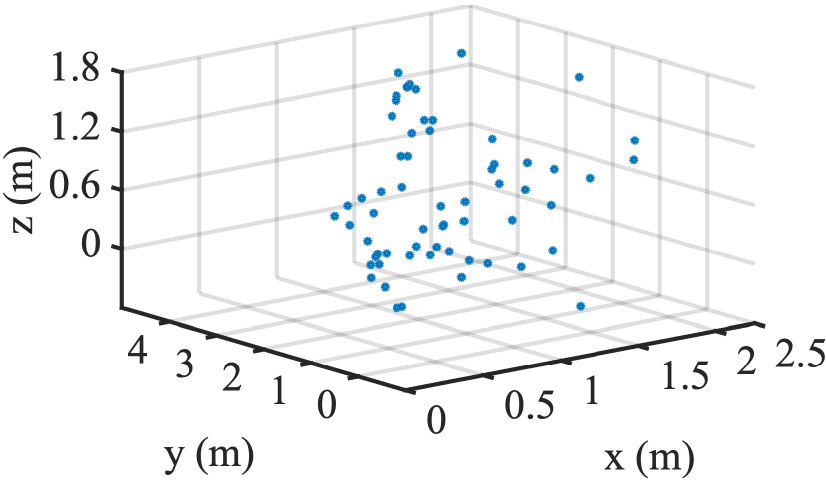}
\label{fig:pointcloudrr}
}
\subfigure[TI radar.]{
\centering
\includegraphics[width=0.22\textwidth]{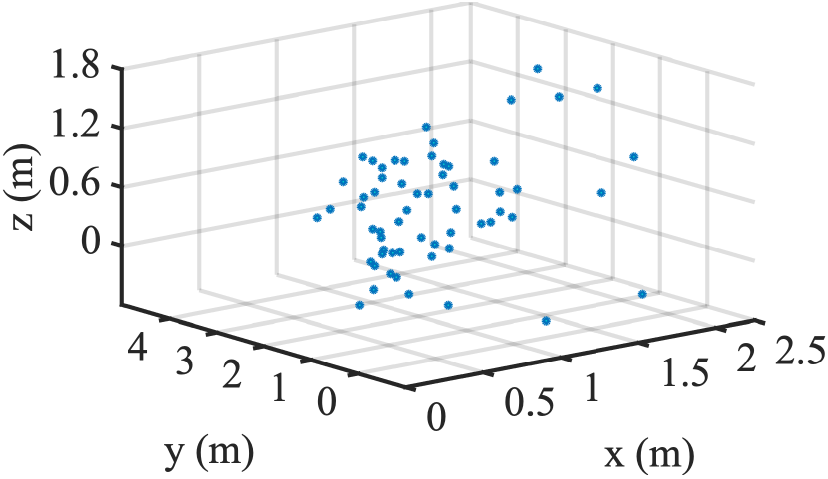}
\label{fig:pointcloudti}
}
\hspace{0.5ex}
\caption{Scheduling for simultaneous communications and point cloud generation (of a human figure).}
\label{fig:point_cloud}
\vspace{-.5ex}
\end{figure}

\vspace{-.5ex}
\subsubsection{Point Cloud} \label{sssec:point_cloud}
%\vspace{-1ex}
% In this experiment, a subject is assigned to stand at front of our mmWave platform, and he/she always sways body while we collect signals for point cloud. 
Different from continuous sensing application in Section~\ref{sssec:respiration}, point cloud is a one-shot sensing application~(e.g., 1\!~s duration in our experiment). In such applications, a much narrower beam is required to generate dense point cloud, yet \name still needs to maintain its beam schedules. Our solution is to have asymmetric Tx and Rx beam patterns for the AP: while the Tx one is designed to match that determined by the corresponding BC-Set,
% into several sectors with a $5^{\circ}$ beamwidth, and scan its sectors in sequence. 
the Rx chain tunes its beam to the finest $1.5^{\circ}$ beamwidth and quickly switches to distinct azimuth and elevation angles for ``scanning'' the subject under the BC-Set coverage. According to the reflection power at different angle pairs, we set a threshold to filter out reflection signals with very lower power, and map the residue signals from polar coordinates to Cartesian coordinates so as to generate point cloud~\cite{gao2019experiments}. 

We test \name and the baselines in sequence and report the results in Figure~\ref{fig:point_cloud}:
% in different environments including lab, corridor, bedroom and meeting room. We calculate 
while throughput is statistics from all trials, the one-shot point cloud results are arbitrarily chosen examples (a human figure) for demonstration purpose only.
%
% to demonstrate in the following. 
% Beam pattern determined by each BC-Set is usually used by Tx chain to cover a BC-Set. 
% Rx chain is utilized to scan azimuth and elevation bearings in a BC-Set coverage for point cloud application. 
% For point cloud application, we need to catch reflection signals with different bearings bounded off a subject. 
% \needrev{Results shown in Figure~\ref{fig:point_cloud} illustrate \name still achieve a comparable throughput with CO, and denser 3D points than CO, RR and TI for point cloud application. }
%[zhe] all throughput except for CO is lower. Why? explain
%
Results in Figure~\ref{fig:pointcloudthroughput} suggest that the throughput gap between \name and CO gets slightly wider, but \name still outperforms other two baselines: to generate 3D point cloud, \name has to temporarily sacrifice a bit throughput to capture fine-grained reflection signals. 
Though SO in Figure~\ref{fig:pointcloudSO} performs marginally better than \name in Figure~\ref{fig:pointcloudGemini}, the main beam of SO has to solely target the subject, hence resulting in the lowest throughput. RR sits in the middle of CO and SO for both throughput and point cloud performance shown in Figure~\ref{fig:pointcloudthroughput} and Figure~\ref{fig:pointcloudrr}, due to its time-division nature. 
Finally,  the TI cascaded radar with a number of antennas~(12 Tx $\times$ 16 Rx) also performs poorly with its lower bearing resolution than Sivers' phased array antennas~\cite{SiversBF01_2021} illustrated in Figure~\ref{fig:pointcloudti}.

% Results of point cloud are plotted from Figure~\ref{fig:pointcloudGemini} to~\ref{fig:pointcloudti} which illustrates \name achieve denser 3D points than CO, RR and TI, but a little less than SO. TI radar does not perform well, due to the limited number of antennas. Figure~\ref{fig:pointcloudthroughput} shows for this application, \needrev{\name still achieve a comparable throughput with CO.}

\begin{figure}[!b]
\vspace{-4ex}
\setlength\abovecaptionskip{8pt}
\subfigure[Throughput~(Gbps).]{
\includegraphics[width=0.22\textwidth]{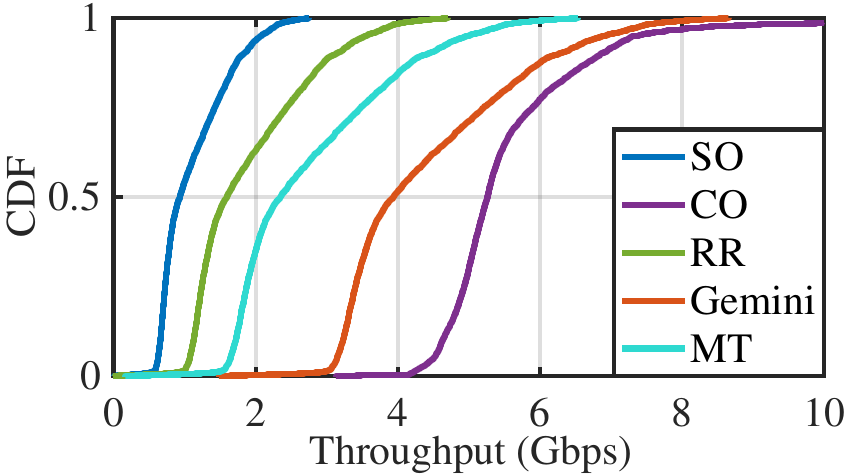}
\label{fig:TrackingThroughput}
}
\hspace{0.4ex}
\subfigure[Tracking RMSE~(m).]{
\includegraphics[width=0.22\textwidth]{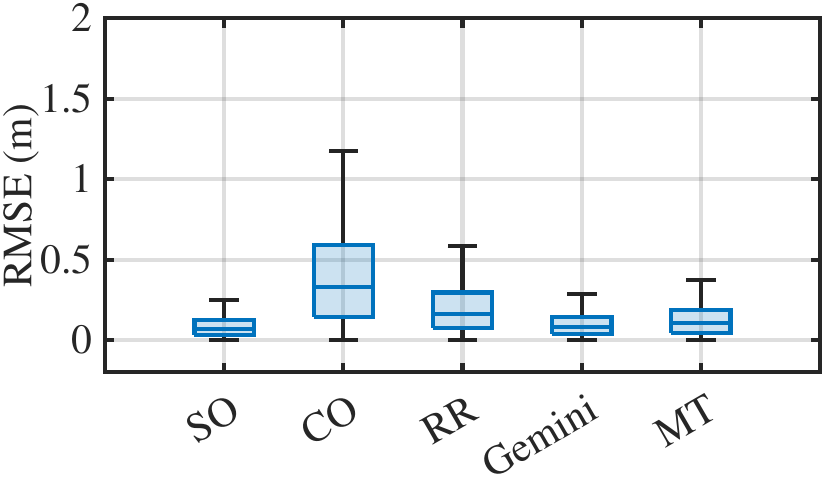}
\label{fig:TrackingRMSE}
}
\vspace{-1ex}
\subfigure[\name-Single target~(Gemini).]{
\centering
\includegraphics[width=0.14\textwidth]{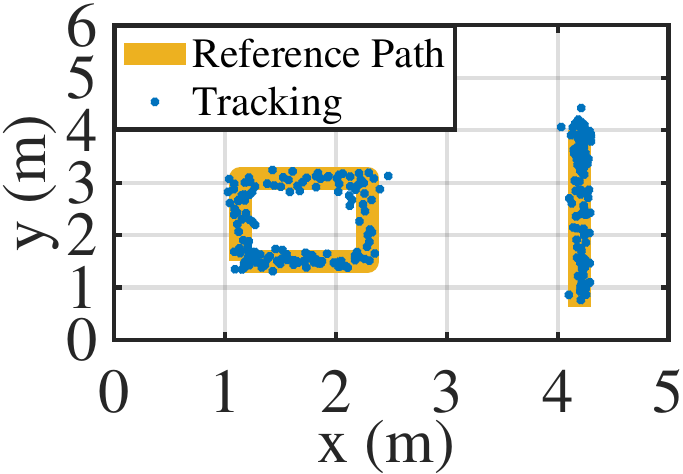}
\label{fig:trackingGemini}
}
\hspace{0.1ex}
\subfigure[Sensing only (SO).]{
\centering
\includegraphics[width=0.14\textwidth]{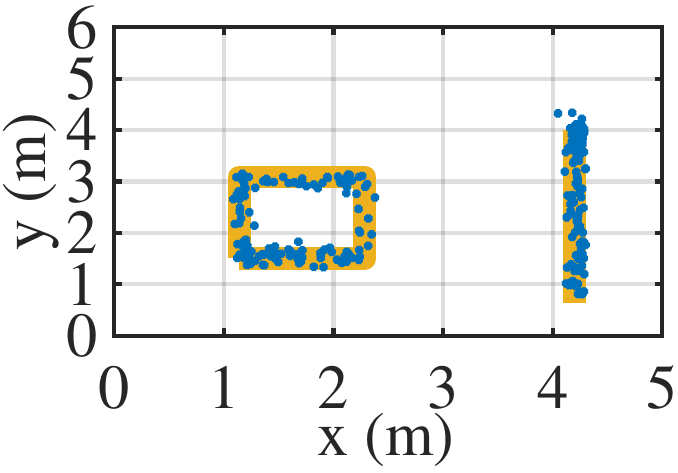}
\label{fig:trackingSO}
}
\hspace{0.1ex}
\subfigure[Comm. only (CO).]{
\centering
\includegraphics[width=0.14\textwidth]{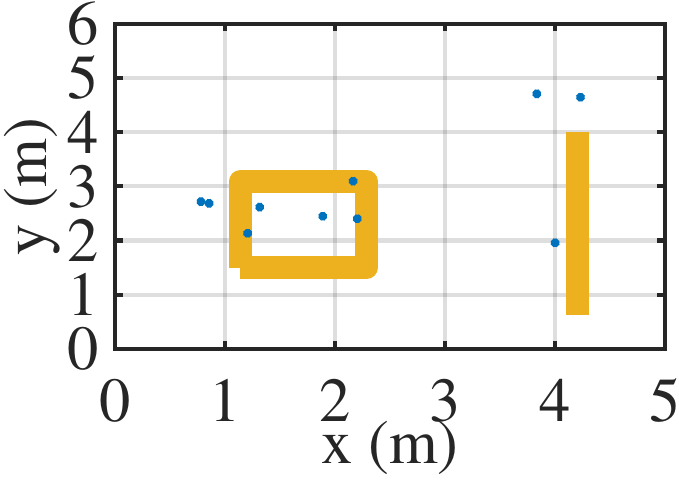}
\label{fig:trackingCO}
}
\vspace{-1ex}
\subfigure[Round-robin (RR).]{
\centering
\includegraphics[width=0.14\textwidth]{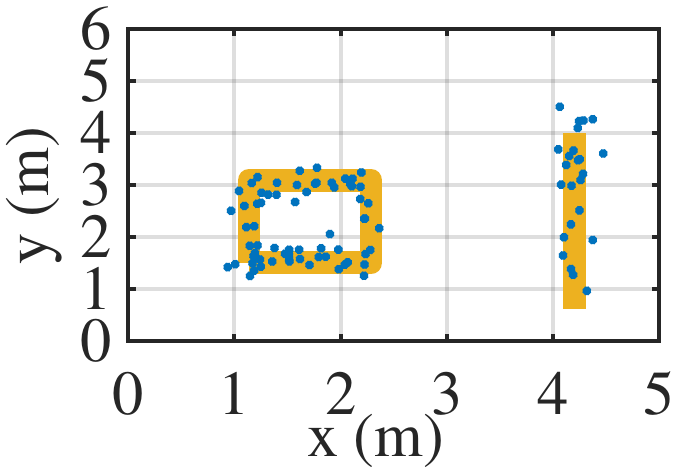}
\label{fig:trackingRR}
}
\hspace{0.1ex}
\subfigure[TI radar~(TI).]{
\centering
\includegraphics[width=0.14\textwidth]{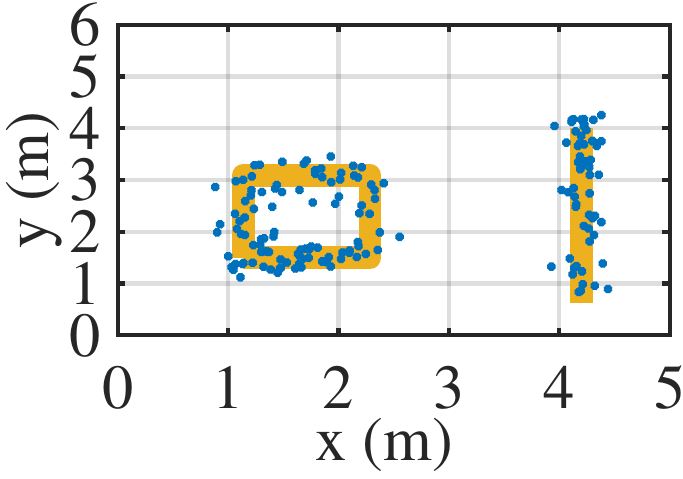}
\label{fig:trackingTI}
}
\hspace{0.1ex}
\subfigure[\name-Multiple targets(MT).]{
\centering
\includegraphics[width=0.14\textwidth]{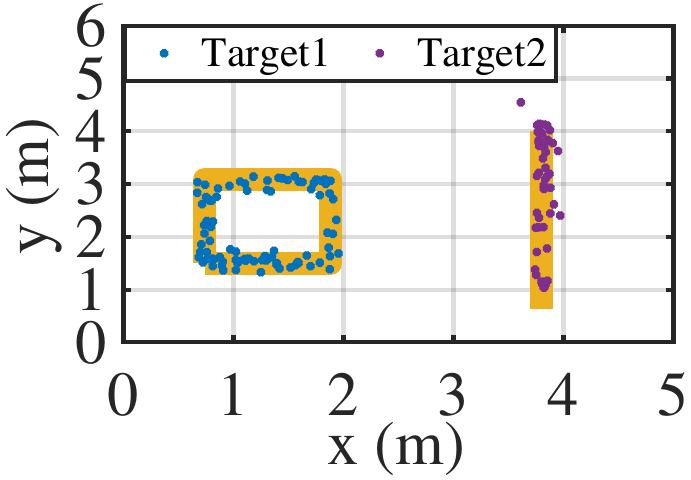}
\label{fig:trackingmultiple}
}
\vspace{-2ex}
\caption{Scheduling for simultaneous communications and human movement tracking in the test area. Orange lines show the ground truth—predefined paths—while blue points display tracking results.}
% \needrev{ same to figure 17, but tracking error bars with 3 words ``Mob''. }
\label{fig:human_tracking}
\vspace{-.5ex}
\end{figure}

\vspace{-0.5ex}
\subsubsection{Human Movement}
\vspace{-0.5ex}
We captured the human movements in a room approximately 21$\times$18 feet, slightly larger than a typical US family room,  accommodating scenarios with larger movements involving multiple UEs and several subjects.
Human subjects follow the predefined paths marked on the floor,  specifically 'Rectangular' and 'Straight Line' trajectories as depicted in Figure~\ref{fig:human_tracking}. Throughput and tracking performance are depicted in Figure~\ref{fig:human_tracking}.
%Gesture tracking sits somewhere in-between continuous and one-shot sensing application, so the same scheduling adopted by Section~\ref{sssec:point_cloud} applies to this experiment. We ask subjects to draw three characters ``M'', ``O'' and ``B'' in the air and track their gesture motions. We plot throughput and tracking performance in Figure~\ref{fig:gesture_tracking} and Table~\ref{table:track}. 
% Apparently, for throughput, \name performs better than SO and RR, but is little worse than CO illustrated in~\ref{fig:TrackingThroughput}.  However, 
%While sharing similar ranking as in Figure~\ref{fig:pointcloudthroughput}, throughput of all schemes shown in Figure~\ref{fig:TrackingThroughput} are slightly lower,
% than those in Figure~\ref{fig:pointcloudthroughput}
%since gesture tracking requires a subject to get close to the AP, hence partially blocking the Tx beam. 
%This phenomenon also leads to similar gesture tracking performance for both \name and SO shown in Figure~\ref{fig:trackingGemini} and Figure~\ref{fig:trackingSO}, since Tx beams generated by both two schemes are easily overlapped for covering whole body of a subject. RR with time-division nature and TI radar with limited spatial resources achieve similar performance illustrated in 
As for the sensing performance shown in Figures~\ref{fig:trackingGemini} to~\ref{fig:trackingTI}, human movement exhibits a similar performance ranking as static point cloud generation, as it is also about producing point cloud but in both temporal and spatial dimensions.
For the case involving multiple human subjects and UEs as shown in Figure~\ref{fig:trackingmultiple}, we placed two frontends as UEs to receive the data from the AP, with each subject following one of the two predefined paths.
Despite the slight decreases in throughput and tracking accuracy observed with two UEs tracking two targets (MT), \name still outperforms baselines with only one UE and one target. It is attributed to our beam scheduling algorithm, which effectively balances communication and sensing tasks.

\iffalse
\begin{table}[t]
\centering
\small
\begin{tabular}{|c|l|c|c|c|c|c|}
\hline
          & Single  &  Multiple               & SO & CO & RR & TI  \\ \hline
RMSE (m) & \multicolumn{1}{c|}{\textbf{0.095}} &  {\textbf{0.124}} & 0.083  &  0.392  & 0.196  & 0.136 \\ \hline
\end{tabular}
% \vspace{1ex}
\caption{Tracking error for human movement.}
\label{table:track}
\vspace{-3em}
\end{table}

\vspace{-0.5ex}
\subsubsection{Remark}
\vspace{-0.5ex}
\fi

%
In fact, the idea of ISAC is not meant to benefit any specific applications; it is basically striking a balance between sensing and communication, \textit{at the benefit of having them co-existing in one hardware}. Meanwhile, \name, with its innovative scheduling driven by BC-Set, is superior to existing mmWave-ISAC solutions, as it is demonstrated by our intensive experiments to have obtained performance very close to either SO (for sensing) or CO (for communications), given that both sensing and communication functions co-exist in one hardware and operate simultaneously. 
% \vspace{-2ex}
% Apparently, similar to Section~\ref{sssec:point_cloud}, except for CO, \name also has the advantage in communication for other baselines.  We find that the performance of \name and SO are almost same, since for gesture tracking, subjects are very close to AP, and the direction of Tx main beams for \name and SO, respectively, overlaps each other. 

% \paragraph{Remark:} Our beam scheduling algorithm balances communication and sensing in three different level sensing applications that is tiny, small and large motion.  
% \vspace{-2ex}
\vspace{-1.5ex}
\subsection{Sensing Diversity}
\vspace{-0.5ex}
We hereby revisit the point cloud application to evaluate our unified estimation framework explained Section~\ref{ssec:hybrid_sensing} for fusing multiple sensing modalities. Basically, we start from the AP monostatic sensing already shown in Figure~\ref{fig:pointcloudGemini} and gradually fuse in more UE (multi-static) sensing results.
% Since we do not have sufficient mmWave platforms as UEs, we leverage time-division manner to move one UE to different locations to emulate multiple UEs for sensing diversity. This setting is reasonable, because we unify channel parameters, such as bearings, range and motion to fuse, but not raw IQ signals. 
% Due to multi-static mode, we only leverages bearings of reflection signals bounded off a subject to fuse. 
As expected, the density of the point cloud increases positively with the number of UEs from Figures~\ref{fig:bcset11} to~\ref{fig:bcset22}, 
%results are shown in Figure~\ref{fig:sensing_diversity} that comparing with Figure~\ref{fig:pointcloudGemini}, 
while the 3D point scattering getting quickly shrunk, rendering the overall human figure clearer step-by-step.
% and more points are centralized in body of subject with increasing the number of UEs. 

\begin{figure}[!b]
\vspace{-2.5ex}
\setlength\abovecaptionskip{8pt}
\subfigure[One UE.]{
\includegraphics[width=0.22\textwidth]{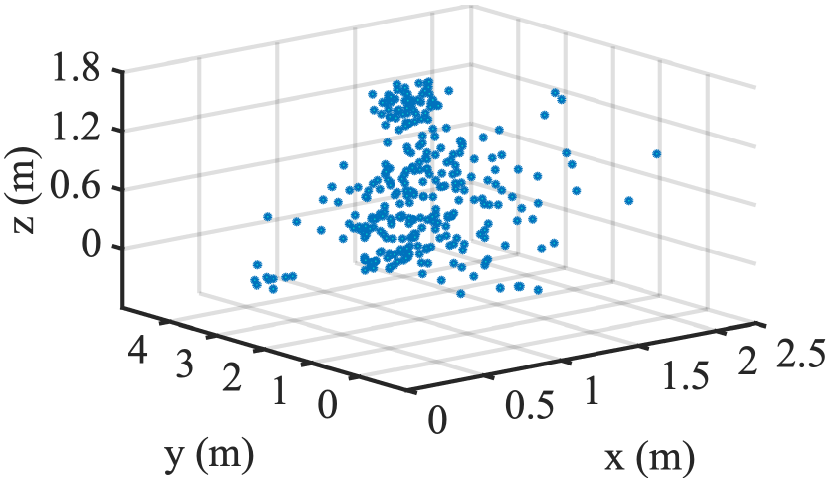}
\label{fig:bcset11}
}
%\hspace{0.5cm}
\subfigure[Two UEs]{
\centering
\includegraphics[width=0.22\textwidth]{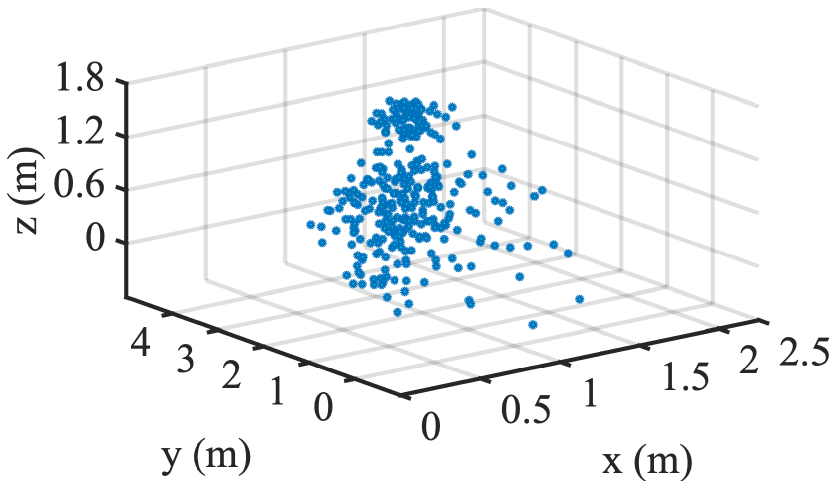}
\label{fig:bcset22}
}
\subfigure[Three UEs]{
\centering
\includegraphics[width=0.22\textwidth]{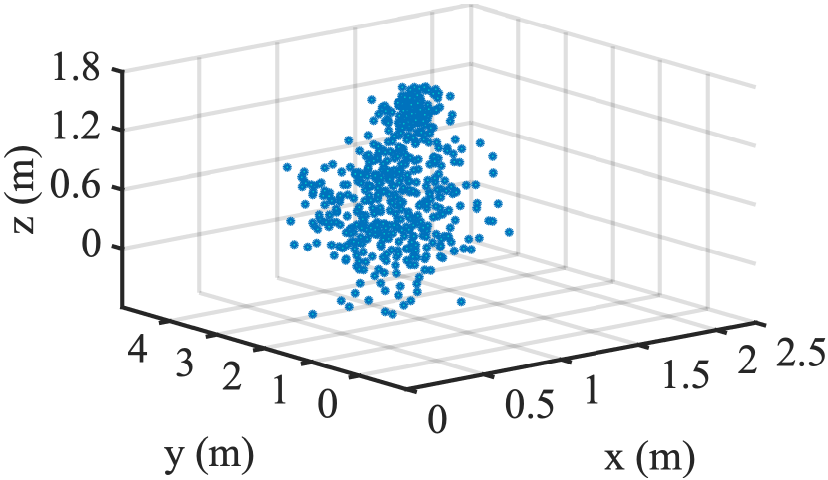}
\label{fig:bcset22}
}
\hspace{0.25ex}
\subfigure[Four UEs]{
\centering
\includegraphics[width=0.22\textwidth]{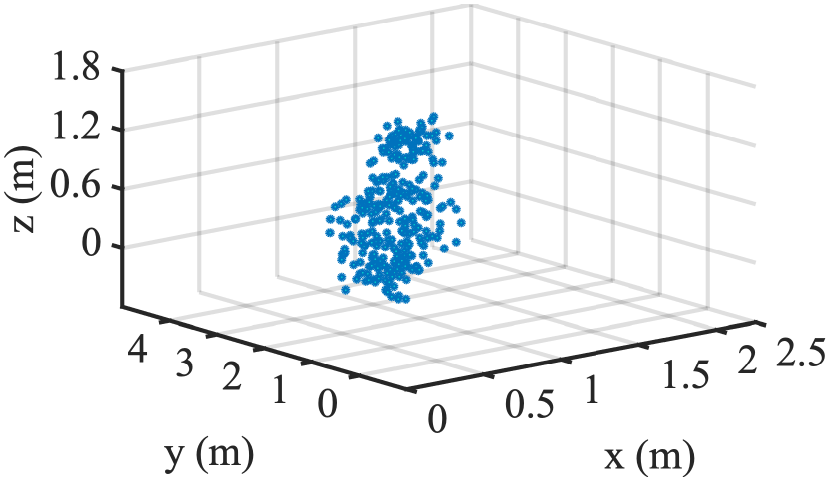}
\label{fig:bcset22}
}
\vspace{-1.5ex}
\caption{Sensing diversity with different UEs may substantially enhance the point cloud quality.}
% \needrev{similar to figure 12, but point cloud is 3D to fuse and demonstrate}
\label{fig:sensing_diversity_e}
\vspace{-.5ex}
\end{figure}

% \section{Discussion} 
% \label{sec:dis}
% \input{discuss}

\vspace{-1ex}
\section{Related Works and Discussions} 
\label{sec:related}
%\vspace{-0.5ex}
%=============== Outlines ================
%1.mmWave Platforms, some cannot support IEEE 802.11ad/ay standards, some don't support bandwidth
%2.mmWave ISAC works
%3.mmWave sensing
%=========================================
%
% Prior to introducing \name, let's give a brief overview of the existing mmWave platforms and joint communication \& sensing systems. With the development of 5G and mm-Wave technology, significant effort has been devoted to creating systems with large bandwidths, smaller components, and concentrated beams at these frequencies to increase performance and enhance sensing accuracy. 
%
% The state-of-the-art literature can be divided into the following categories:

This section focuses on three main streams of related works, namely mmWave platforms, ISAC, and full duplex radio~(FDR), with brief discussions on limitations of \name.
%
% Existing proposals on nulling for communications~\cite{RobertsFDmmWave,mmWaveFD-MobiCom20,nullifi-NSDI21} are only marginally related and hence omitted from our discussion.
% Our \name is learned from earlier mmWave communication research works, but they mainly focus on nulling Tx interference for communication.
% Fundamental difference between full duplex and ISAC is been analyzed in~\cite{isacot}. 

\vspace{-1ex}
\paragraph{mmWave Platforms} Earlier mmWave platforms~\cite{openmili-MobiCom16, X60-WiNTECH17} only offer a relatively wide bandwidth, yet their phased arrays cannot be precisely controlled for fast alignment, thus not exactly meeting the requirements of IEEE 802.11ad~\cite{802_11ad}. M-Cube~\cite{MCube-MobiCom20} utilizes a commodity 802.11ad RF front-end with phased array antennas,
% which are connected to an FPGA board for processing data. 
but it was soon superseded by mmFlex~\cite{mmFLEX-MobiSys20} adopting similar mmWave front-end as \name to also match the more advanced standard of 802.11ay.
% a fast real-time phased array reconfiguration scheme, and above 2\!~GHz bandwidth that meet both IEEE 802.11ad and IEEE 802.11ay standard requirements. Our \name is build on similar RF front-end of mmFlex to develop a novel mmWave ISAC system. 

% \needrev{wivi and wideo are missing}

\vspace{-1.5ex}
\paragraph{ISAC} 
Though theoretical works studying mmWave ISAC for 5/6G cellular network scenarios are plentiful~\cite{liu2018toward, ZhangJCS-CommSurvey21}, they only focus on waveform design without providing useful guidelines for system developments. 
% In order to bridge gap between theory and practice, 
For ISAC on mmWave band, a couple of proposals~\cite{SPARCS-IPSN22, SideLobe-UbiComp23} have leveraged existing communication traffic to enable only multi-static sensing confined to subjects compatible with existing communication beam patterns. Though a monostatic sensing solution has been mentioned in~\cite{guan-TMTT2021, JUMP_TWC24}, it is far from fully integration into existing communication devices. 
%it is not clear how it can be integrated into a communication device.
%\needrev{However, above system works are special design for only sensing, but incompatible with current commodity mmWave devices leading to lower throughput~(explained in Section~\ref{ssec:beam_schedule}). }

% A few works, such as~\cite{SPARCS-IPSN22,SideLobe-UbiComp23}  leverage existing communication packets to enable multi-static sensing, and only work~\cite{guan-TMTT2021} has proposed a monostatic sensing solution based on SDR. However, above works are not a full-fledged ISAC system. 

\vspace{-1.5ex}
\paragraph{Nulling and FDR}
Existing proposals on nulling for communications~\cite{RobertsFDmmWave,mmWaveFD-MobiCom20,nullifi-NSDI21} are only marginally related and hence omitted from our discussions. Certain Wi-Fi sensing developments~\cite{adib2013see,joshi2015wideo} have gone very close to FDR~\cite{FDR-SIGCOMM13}, as they adopt either nulling or FDR to cancel Tx interference for motion sensing. However, as pointed out by~\cite{isacot}, monostatic sensing for ISAC is fundamentally different from FDR. 

\vspace{-1.5ex}
\paragraph{\name} 
\name cannot leverage FDR-like technologies to extract monostatic sensing signals~(explained in Section~\ref{ssec:mono_sens}), so we exploit a hybrid (hardware-software) deep learning model to remove Tx interference.
% but is still comparable with state-of-the-art mmWave communication architecture. 
We also innovate in designing a beam scheduling for ISAC~(see Section~\ref{ssec:sche_isac}), and fusing multiple sensing modalities for improving estimation precision~(see Section~\ref{ssec:hybrid_sensing}). 
%as well as maintaining full compatibility with 802.11ay~(see Section~\ref{ssec:hybrid_sensing}).}
% ~(see Section~\ref{ssec:protocol_comp}). 
In the meantime, we are considering fusing sub-6\!~GHz Wi-Fi and mmWave to enhance the capability of ISAC, as it has been done for communication only~\cite{sur2017wifi}. Also, endowing Wi-Fi with sensing capability may cause unexpected information leakage~\cite{li2016csi}, which can be exacerbated by the powerful mmWave-ISAC, so security issue should be part of our future work.

\vspace{-.5ex}
\section{Conclusion} 
\label{sec:con}
%\vspace{-.5ex}

In this paper, we have proposed, designed and implemented \name, a full-fledge mmWave ISAC system. We have first given three concrete analyses to motivate our design. then we have elaborated all key components for \name, namely i) Tx interference cancellation driven by deep learning to enable monostatic sensing, ii) beam scheduling algorithm for jointly optimizing sensing accuracy and communication throughput, and iii) a unified estimation framework for exploiting diversified sensing modalities. Finally, we have conducted extensive experiments to evaluate the performance of \name; our results have strongly demonstrated advantages of \name in actually taking care of both communication and sensing under its ISAC framework.
We believe that our initial trials in realizing \name 
% comprehensive range of studies 
signify a pivotal step towards more practical mmWave ISAC systems.

\balance
\bibliographystyle{abbrv}
% \small

\bibliography{reference}

\clearpage
\normalsize
\appendix

\end{document}